\documentclass[iop,revtex4.1]{emulateapj}

\usepackage{natbib}
\usepackage{amsmath}
\usepackage{color}
\usepackage{url}
\usepackage{ulem}
\usepackage{multirow}
\usepackage{amsmath}
\usepackage{wasysym}

\bibpunct{(}{)}{;}{a}{}{,}

\def\apjl{ApJ}%
\def\apjs{ApJS}%
\def\ao{Appl.~Opt.}%
\def\aap{A\&A}%

\shorttitle{ IC\,751: a new changing-look AGN discovered by {\it NuSTAR}}
\shortauthors{Ricci et al.}

\begin{document}


\title{IC\,751: a new changing-look AGN discovered by {\it NuSTAR}

}


%

\author{C. Ricci\altaffilmark{1,2*}, F. E. Bauer\altaffilmark{1,2,3,4}, P. Arevalo\altaffilmark{5},  S. Boggs\altaffilmark{6}, W. N. Brandt\altaffilmark{7,8,9}, F. E. Christensen\altaffilmark{10}, W. W. Craig\altaffilmark{6}, P. Gandhi\altaffilmark{11},  C. J. Hailey\altaffilmark{12}, F. A. Harrison\altaffilmark{13}, M. Koss\altaffilmark{14}, C. B. Markwardt\altaffilmark{15,16}, D. Stern\altaffilmark{17}, E. Treister\altaffilmark{18,2}, W. W. Zhang\altaffilmark{16}
}

\altaffiltext{1}{Instituto de Astrof\'{\i}sica, Facultad de F\'{i}sica, Pontificia Universidad Cat\'{o}lica de Chile, Casilla 306, Santiago 22, Chile} 
\altaffiltext{2}{EMBIGGEN Anillo}
\altaffiltext{3}{Millennium Institute of Astrophysics, Vicu\~{n}a Mackenna 4860, 7820436 Macul, Santiago, Chile} 
\altaffiltext{4}{Space Science Institute, 4750 Walnut Street, Suite 205, Boulder, Colorado 80301, USA}
\altaffiltext{5}{Instituto de F\'isica y Astronom\'ia, Facultad de Ciencias, Universidad de Valpara\'iso, Gran Bretana N¼ 1111, Playa Ancha, Valpara\'iso, Chile.}
\altaffiltext{6}{ Space Sciences Laboratory, University of California, Berkeley, CA 94720, USA}
\altaffiltext{7}{Department of Astronomy and Astrophysics, The Pennsylvania State University, University Park, PA 16802, USA}
\altaffiltext{8}{Institute for Gravitation and the Cosmos, The Pennsylvania State University, University Park, PA 16802, USA}
\altaffiltext{9}{Department of Physics, 104 Davey Lab, The Pennsylvania State University, University Park, PA 16802, USA}
\altaffiltext{10}{DTU Space, National Space Institute, Technical University of Denmark, Elektronvej 327, DK-2800 Lyngby, Denmark}
\altaffiltext{11}{School of Physics \& Astronomy, University of Southampton, Highfield, Southampton, SO17 1BJ}
\altaffiltext{12}{Columbia Astrophysics Laboratory, Columbia University, New York 10027, USA}
\altaffiltext{13}{Cahill Center for Astronomy and Astrophysics, California Institute of Technology, Pasadena, CA 91125, USA }
\altaffiltext{14}{Institute for Astronomy, Department of Physics, ETH Zurich, Wolfgang-Pauli-Strasse 27, CH-8093 Zurich, Switzerland}
\altaffiltext{15}{Department of Astronomy, University of Maryland, College Park, MD 20742, USA}
\altaffiltext{16}{Astroparticle Physics Laboratory, Mail Code 661, NASA Goddard Space Flight Center, Greenbelt, MD 20771, USA}
\altaffiltext{17}{Jet Propulsion Laboratory, California Institute of Technology, Pasadena, CA 91109, USA}
\altaffiltext{18}{Universidad de Concepci\'on, Departamento de Astronom\'ia, Casilla 160-C, Concepci\'on, Chile}

\altaffiltext{*}{cricci@astro.puc.cl}

\begin{abstract}
We present the results of five {\it NuSTAR} observations of the type\,2 active galactic nucleus (AGN) in IC\,751, three of which were performed simultaneously with {\it XMM-Newton} or {\it Swift}/XRT. We find that the nuclear X-ray source underwent a clear transition from a Compton-thick ($N_{\rm\,H}\simeq 2\times 10^{24}\rm\,cm^{-2}$) to a Compton-thin ($N_{\rm\,H}\simeq 4\times 10^{23}\rm\,cm^{-2}$) state on timescales of $\lesssim 3$ months, which makes IC\,751 the first changing-look AGN discovered by {\it NuSTAR}. Changes of the line-of-sight column density at a $\sim2\sigma$ level are also found on a time-scale of $\sim 48$\,hours ($\Delta N_{\rm\,H}\sim 10^{23}\rm\,cm^{-2}$). From the lack of spectral variability on timescales of $\sim 100$\,ks we infer that the varying absorber is located beyond the emission-weighted average radius of the broad-line region, and could therefore be related either to the external part of the broad-line region or a clumpy molecular torus. By adopting a physical torus X-ray spectral model, we are able to disentangle the column density of the non-varying absorber ($N_{\rm\,H}\sim 3.8\times 10^{23}\rm\,cm^{-2}$) from that of the varying clouds [$N_{\rm\,H}\sim(1-150)\times10^{22}\rm\,cm^{-2}$], and to constrain that of the material responsible for the reprocessed X-ray radiation ($N_{\rm\,H} \sim 6 \times 10^{24}\rm\,cm^{-2}$). We find evidence of significant intrinsic X-ray variability, with the flux varying by a factor of five on timescales of a few months in the 2--10 and 10--50\,keV band. 

\end{abstract}



\section{Introduction}

Variability of the line-of-sight column density ($N_{\rm H}$) might be rather common in Active Galactic Nuclei (AGN) \citep{Risaliti:2002uq,Bianchi:2012cr,Torricelli-Ciamponi:2014dq}, and in the past decade several objects have been found to show eclipses of the X-ray source, both due to Compton-thick (CT, $N_{\rm\,H}\gtrsim 10^{24}\rm\,cm^{-2}$) and to Compton-thin ($N_{\rm\,H}< 10^{24}\rm\,cm^{-2}$) material. Since the X-ray source is believed to be located very close to the supermassive black hole (SMBH), this variable absorption could be associated either with broad-line region (BLR) clouds, or with clumps in the molecular torus.  In at least a few cases (e.g., \citealp{Risaliti:2009mi,Maiolino:2010fu}) these variations have been found to happen on timescales of days, and are consistent with being related to material in the BLR. \cite{Markowitz:2014oq} have recently shown, by studying {\it RXTE} light-curves of 55 AGN, that for eight objects of their sample there seems to be variation of absorbing material also on longer timescales (months to years). They associated these changes in $N_{\rm\,H}$ with clumps in the molecular torus. Interestingly, none of the eclipses detected by \cite{Markowitz:2014oq} were due to CT material.

\begin{table*}
\begin{center}
\caption[]{X-ray observations log.}
\label{tab:obslog}
\begin{tabular}{clccc}
\noalign{\smallskip}
\hline \hline \noalign{\smallskip}
Obs. \#& Facility & Observation date & Observation ID &Net Exposure [ks]\\
\noalign{\smallskip}
\hline \noalign{\smallskip}
1  & {\it Swift}/XRT$^{a}$ 	 												& 2008-02-20 13:13:01	& 00037374001	& 2.3		\\
\noalign{\smallskip}
\noalign{\smallskip}
2 &   {\it NuSTAR}													&  2012-10-28 23:01:07	& 60061217002	& 13.1		\\ 
\noalign{\smallskip}
\noalign{\smallskip}
 3 &  {\it NuSTAR}													& 2013-02-04 00:26:07  	&  60061217004 	& 52.0		\\ 
\noalign{\smallskip}
\noalign{\smallskip}
4 &   {\it NuSTAR}													& 2013-05-23 05:36:07	 & 60061217006 	& 25.0		\\ 
4 &   {\it Swift}/XRT 	 												&  2013-05-25 16:38:59	& 00080064001	& 5.8		\\  
\noalign{\smallskip}
\noalign{\smallskip}
 5 &  {\it NuSTAR}													& 2014-11-28 06:01:07  	&  60001148002	& 26.3		\\ 
 5 &  {\it XMM-Newton} 	 											& 2014-11-28 13:20:42		& 0744040301 &	18.4$^{b}$; 23.1$^{c}$		\\ 
\noalign{\smallskip}
\noalign{\smallskip}
 6 &  {\it NuSTAR}													& 2014-11-30 06:26:07  	&  60001148004 	& 25.7		\\ 
 6 &  {\it XMM-Newton} 	 											& 2014-11-30 13:11:46	 	& 0744040401	& 	18.2$^{b}$; 22.4$^{c}$	\\
\noalign{\smallskip}
\noalign{\smallskip}
\hline
\noalign{\smallskip}
\multicolumn{5}{l}{{\bf Notes}. $^{a}$ this observation was not used for spectral fitting because of the}\\
\multicolumn{5}{l}{ low number of counts;  $^{b}$ EPIC/PN; $^{c}$ EPIC MOS1 \& MOS2}\\
\end{tabular}

\end{center}
\end{table*}

So far variations in the $N_{\rm H}$ of the neutral absorber have been found in more than twenty AGN including 1H\,0419$-$577 \citep{Pounds:2004bh}, Centaurus\,A \citep{Beckmann:2011hc,Rivers:2011kl}, ESO\,323$-$G77 \citep{Miniutti:2014nx}, H0557-385 \citep{Longinotti:2009tg}, MR 2251-178, Mrk\,348, Mrk\,509 \citep{Markowitz:2014oq}, Mrk\,6 \citep{Immler:2003fk}, Mrk\,766 \citep{Risaliti:2011fk}, Mrk\,79 \citep{Markowitz:2014oq}, NGC\,1068 \citep{Marinucci:2015kx}, NGC\,1365 \citep{Risaliti:2005kl,Risaliti:2007qa,Maiolino:2010fu,Walton:2014fk,Rivers:2015uq}, NGC\,3227 \citep{Lamer:2003oq}, NGC\,3783 \citep{Markowitz:2014oq}, NGC\,4151 \citep{Puccetti:2007zr}, NGC\,4388 \citep{Elvis:2004vn}, NGC\,4395 \citep{Nardini:2011ij}, NGC\,4507 \citep{Braito:2013nx,Marinucci:2013pi}, NGC\,454 \citep{Marchese:2012kx}, NGC\,5506 \citep{Markowitz:2014oq}, NGC\,6300 \citep{Guainazzi:2002dz}, NGC\,7582 (\citealp{Piconcelli:2007bs,Bianchi:2009ly,Rivers:2015fk}), NGC\,7674 \citep{Bianchi:2005ys}, PG\,2112+059 \citep{Gallagher:2004uq}, UGC\,4203 \citep{Guainazzi:2002fv,Risaliti:2010ve}, and SWIFT\,J2127.4+5654 \citep{Sanfrutos:2013cr}. In most cases these variations are due to Compton-thin material, and only for a handful of sources is the varying absorber CT (i.e., ESO\,323$-$G77, NGC\,1068, NGC\,1365, NGC\,454, NGC\,6300, NGC\,7582, NGC\,7674, UGC\,4203). AGN switching between Compton-thin and CT states are usually dubbed {\it changing-look} AGN (e.g., \citealp{Matt:2003vn}), because their spectral shape changes dramatically (from transmission-dominated to reflection-dominated). In the optical band changing-look AGN are objects that transition from type-1 to type-1.8/1.9/2 (e.g., \citealp{LaMassa:2015ys}), or from type-1.8/1.9/2 to type-1 (e.g., \citealp{Shappee:2014kx}). In the following we will refer to the X-ray classification only.

IC\,751 ($z=0.0311$, $D=137$\,Mpc, \citealp{Falco:1999fk}) is a type\,2 AGN \citep{Veron-Cetty:2010ly} in an edge-on spiral galaxy \citep{de-Vaucouleurs:1991nx} that has not yet been studied in detail in the X-ray band. The source was reported in the 70-month {\it Swift}/BAT catalogue \citep{Baumgartner:2013uq}, and was observed by {\it NuSTAR} as part of the campaign aimed at following-up {\it Swift}/BAT detected sources (Balokovic et al., in prep.). The interesting X-ray characteristics of this object triggered several follow-up observations with {\it NuSTAR}. We report here on the five {\it NuSTAR} observations of this source carried out between 2012 and 2014, two of which were performed jointly with {\it XMM-Newton}. The source switches from a CT to Compton-thin state between the different observations, and is the first changing-look AGN discovered by {\it NuSTAR}. Following our X-ray spectral and temporal analysis we put constraints on the location of the varying obscuring material. Throughout the paper we consider a luminosity distance of the source of $d_{\rm\,L}=135$\,Mpc, and adopt standard cosmological parameters ($H_{0}=70\rm\,km\,s^{-1}\,Mpc^{-1}$, $\Omega_{\mathrm{m}}=0.3$, $\Omega_{\Lambda}=0.7$). Unless otherwise stated, all uncertainties are quoted at the 90\% confidence level.

\section{X-ray observations and data reduction}
IC\,751 was observed five times by {\it NuSTAR}, twice jointly with {\it XMM-Newton} (PI F. Bauer), and two times by {\it Swift}/XRT. Details about these observations are reported in Table\,\ref{tab:obslog}.

\medskip

\subsection{{\it NuSTAR}}
The {\it Nuclear Spectroscopic Telescope Array} ({\it NuSTAR}, \citealp{Harrison:2013uq}) is the first focusing X-ray telescope in orbit operating above 10\,keV. {\it NuSTAR} consists of two focal-plane modules (FPMA and FPMB), both operating in the 3--79\,keV band and with similar characteristics.

{\it NuSTAR} observed IC\,751 five times between October 2012 and November 2014, with exposure times ranging between 13 and 52\,ks (Table\,\ref{tab:obslog}). The data collected by {\it NuSTAR} were processed using the {\it NuSTAR} Data Analysis Software \textsc{nustardas}\,v1.4.1 within Heasoft\,v6.16, using the latest calibration files, released in March 2015 \citep{Madsen:2015uq}.
The source spectra and light-curves were extracted using the \textsc{nuproducts} task, selecting circular regions with a radius of 50$''$. The background spectra and light-curves were obtained in a similar fashion, using a circular region of 60$''$ radius located where no other source was detected. The source light-curve was corrected for background using the \textsc{lcmath} task.

\begin{table*}
\begin{center}
\caption[]{X-ray spectral analysis -- slab model}
\label{tab:resultsFit_XMM_Nustar}
\begin{tabular}{lcccccc}
\noalign{\smallskip}
\hline \hline \noalign{\smallskip}
\multicolumn{1}{l}{\,\,\,\,\,\,\,\,\,\,\,\,\,\,\,\,\,\,\,\,\,\,\,\,\,(1)} & (2) &  (3) &  (4) &  (5) &  (6)    \\
\noalign{\smallskip}
Model Parameters 																& Obs. 2  					& Obs. 3  					& Obs. 4 					& Obs. 5 								& Obs. 6   								\\
\noalign{\smallskip}
	 																			& [2012-10-28] 				& [2013-02-04]				& [2013-05-23]  			& [2014-11-28] 							& [2014-11-30] 							\\
\noalign{\smallskip}
\hline \noalign{\smallskip}
Photon index $\Gamma$	 														& $1.96^*$  				& $1.74^*$	 				& $1.73^*$					& 	$1.93^{+0.13}_{-0.12}$				& 	  $1.89^{+0.17}_{-0.17}$  	  		\\
\noalign{\smallskip}
Cutoff energy $E_{\rm\,C}$	 	[{\tiny keV}] 									& $\geq 222^*$				& $\geq 222^*$				&  $\geq 222^*$				& 	$\geq 192$							& 	  $\geq 222$  						\\
\noalign{\smallskip}
Reflection param. $R$ 	 														& $0.02^*$ 					& $0.27^*$ 					& $0.28^*$ 					& 	$\leq 0.17$		 					& 	 $0.14^{+0.16}_{-0.13}$	  			\\
\noalign{\smallskip}
Col. density $N_{\rm H}$ [{\tiny $10^{22}\rm\,cm^{-2}$}] 	 					& $199^{+75}_{-67}$			& $121^{+29}_{-27} $		&	$32^{+9}_{-6}$			& 	$38^{+3}_{-3}$		 				& 	$49^{+7}_{-6}$  	 				\\
\noalign{\smallskip}
Fraction of scattered component $f_{\rm\,scatt}$ 	 [{\tiny $\%$}]				&  $17.3^{+9.0}_{-10.5}$ 	& $0.9^*$	 	& $0.9^*$		 			& 	$0.5^{+0.2}_{-0.2}$	 				&  	$0.6^{+0.3}_{-0.2}$	 				\\
\noalign{\smallskip}
Temperature $kT$		[{\tiny keV}] 	 	 									& $0.94^{\dagger}$ 					& $0.94^{\dagger}$ 					& $1.01^*$					& 	$0.82^{+0.15}_{-0.09}$		 		& 	 $0.94^{+0.07}_{-0.19}$	  			\\
\noalign{\smallskip}
Energy (Fe\,K$\alpha$)	[{\tiny keV}] 											& $6.25^*$ 					& $6.36^*$ 					& $6.25^*$					& 	$ 6.42_{-0.09}^{+0.08}$		 		& 	$6.36_{-0.11}^{+0.10}$	 			\\
\noalign{\smallskip}
Norm. (Fe\,K$\alpha$)	[{\tiny $10^{-6}$ photons $\rm\,cm^{-2} \,s^{-1}$}] 	& $2.1^*$ 					& $1.9^*$ 					& $1.5^*$ 					& 	$ 1.4_{-1.0}^{+1.1}$		 		& 	$1.3_{-0.8}^{+0.8}$	  				\\
\noalign{\smallskip}
EW(Fe\,K$\alpha$)	[{\tiny eV}] 												& $281^{+43}_{-218}$					& $489_{-26}^{+955}$			& $130^{+617}_{-65}$			& 	$ 60_{-44}^{+33}$		 			& 	 $108_{-39}^{+64}$	 				\\
\noalign{\smallskip}
$F_{\,2-10}^{\rm\,obs}$ 	[{\tiny $10^{-12}\,\rm\,erg\,cm^{-2}\,s^{-1}$}]				& $0.7^{+0.1}_{-0.5}$						&  $0.22^{+0.01}_{-0.08}$					& 	$0.62^{+0.02}_{-0.06}$					&	$1.11^{+0.07}_{-0.12}$							&	$0.56^{+0.03}_{-0.09}$		 						\\
\noalign{\smallskip}
$F_{\,2-10}$ [{\tiny $10^{-12}\rm\,erg\,cm^{-2}\,s^{-1}$}]							& $22^{+3}_{-15}$							& $3.4^{+0.2}_{-1.2}$ 					& 	$2.1^{+0.1}_{-0.2}$						&	$5.9^{+0.4}_{-0.6}$								&	$3.8^{+0.2}_{-0.6}$		 						\\
\noalign{\smallskip}
$F_{\,10-50}^{\rm\,obs}$ 	[{\tiny $10^{-12}\,\rm\,erg\,cm^{-2}\,s^{-1}$}]				& $4.2^{+0.5}_{-2.4}$						& $3.5^{+0.1}_{-1.6}$					& 	$3.9^{+0.3}_{-0.3}$						&	$5.2^{+0.3}_{-0.6}$								& 	$3.5^{+0.2}_{-0.7}$		  						\\
\noalign{\smallskip}
$F_{\,10-50}$ [{\tiny $10^{-12}\,\rm\,erg\,cm^{-2}\,s^{-1}$}]							& $22^{+3}_{-13}$							&  $6.7^{+0.2}_{-3.1}$					& 	$4.7^{+0.4}_{-0.4}$						&	$7.3^{+0.4}_{-0.8}$								&	$5.2^{+0.3}_{-1.0}$		 						\\
\noalign{\smallskip}
$\log L_{\,2-10}$ [{\tiny $\rm\,erg\,s^{-1}$}]										& 43.7									& $42.9$								& 	$42.7$								&	$43.1$										&	$42.9$		 						\\
\noalign{\smallskip}
$\log L_{\,10-50}$ [{\tiny $\,\rm\,erg\,s^{-1}$}]									& 43.7									& $43.2$ 								& 	$43.0$								&	$43.2$										&	$43.1$		 						\\
\noalign{\smallskip}
\hline
\noalign{\smallskip}
$\chi^2/\mathrm{DOF}$	 												&	31.3/27				 				& 84.7/95 					& 	131.9/111$^{\rm A}$		& 	256.8/257		 					& 224.8/176		 						\\
\noalign{\smallskip}
\hline
\noalign{\smallskip}
\end{tabular}
\tablecomments{The table reports the model parameters obtained by fitting the X-ray spectra of the five epochs with the slab model: \textsc{constant$\times$tbabs$_{\rm\,Gal}\times$(ztbabs$\times$cabs$\times$cutoffpl + apec + pexrav + zgauss + constant$\times$cutoffpl)} in XSPEC; The luminosities reported here are the intrinsic (i.e., absorption-corrected) values. For details see Sect.\,\ref{sect:Xrayspecanal1}. $^*$ value of the parameter left free to vary within the uncertainties of Obs.\,6. $^\dagger$ value of the parameter fixed. $^{\rm A}$ Poissonian statistics applied for {\it Swift}/XRT data. The uncertainties reported for the Fe K$\alpha$ EW correspond to the $68\%$ confidence interval.}

\end{center}
\end{table*}

\subsection{{\it XMM-Newton}}
The {\it XMM-Newton} X-ray observatory \citep{Jansen:2001vn} observed IC\,751 twice at the end of November\,2014. We analysed the two $\sim 20$\,ks {\it XMM-Newton} observations of IC\,751, taking into account the data obtained by the PN \citep{Struder:2001uq} and MOS \citep{Turner:2001fk} cameras. The observation data files (ODFs) were reduced using the the {\it XMM-Newton} Standard Analysis Software (SAS) version 12.0.1 \citep{Gabriel:2004fk}. The raw PN (MOS) data files were then processed using the \textsc{epchain} (\textsc{emchain}) task.

For both observations we analyzed the background light curves in the 10--12 keV band (EPIC/PN), and above 10\,keV (EPIC/MOS), to filter the exposures for periods of high background activity. We set the threshold to $0.5\rm\,ct\,s^{-1}$ and to $0.3\rm\,ct\,s^{-1}$ for PN and MOS, respectively. This resulted in about $10\%$ of the observations being filtered out. We report the final exposures used in Table\,\ref{tab:obslog}. Only patterns that correspond to single and double events (PATTERN~$\leq 4$) were selected for PN, and corresponding to single, double, triple and quadruple events for MOS (PATTERN~$\leq 12$). 

For the three cameras, the source spectra were extracted from the final filtered event list using circular regions centred on the object, with a radius of 20$''$, while the background was estimated using circular regions with a radius of 40$''$ located on the same CCD as the source, where no other source was present. No pile-up was detected for any of the three cameras in the two observations. The ARFs and RMFs were created using the \textsc{arfgen} and \textsc{rmfgen} tasks, respectively. For both observations the source and background spectra of the two MOS cameras, together with the RMF and ARF files, were merged using the \textsc{addascaspec} task.

\begin{figure*}[H!]
\centering
\begin{minipage}[!b]{.48\textwidth}
\centering
\includegraphics[width=8cm]{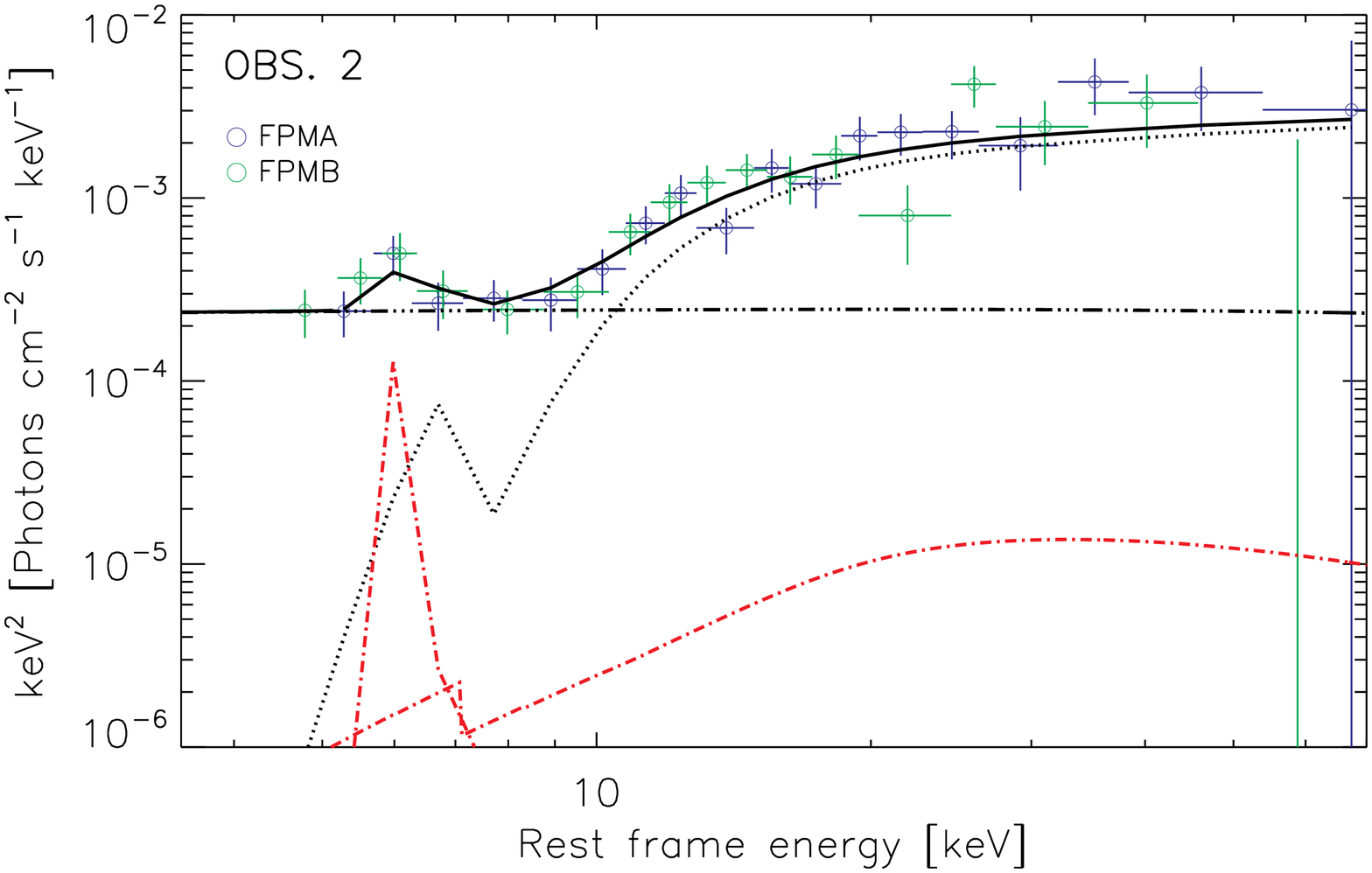}\end{minipage}
\begin{minipage}[!b]{.48\textwidth}
\centering
\includegraphics[width=8cm]{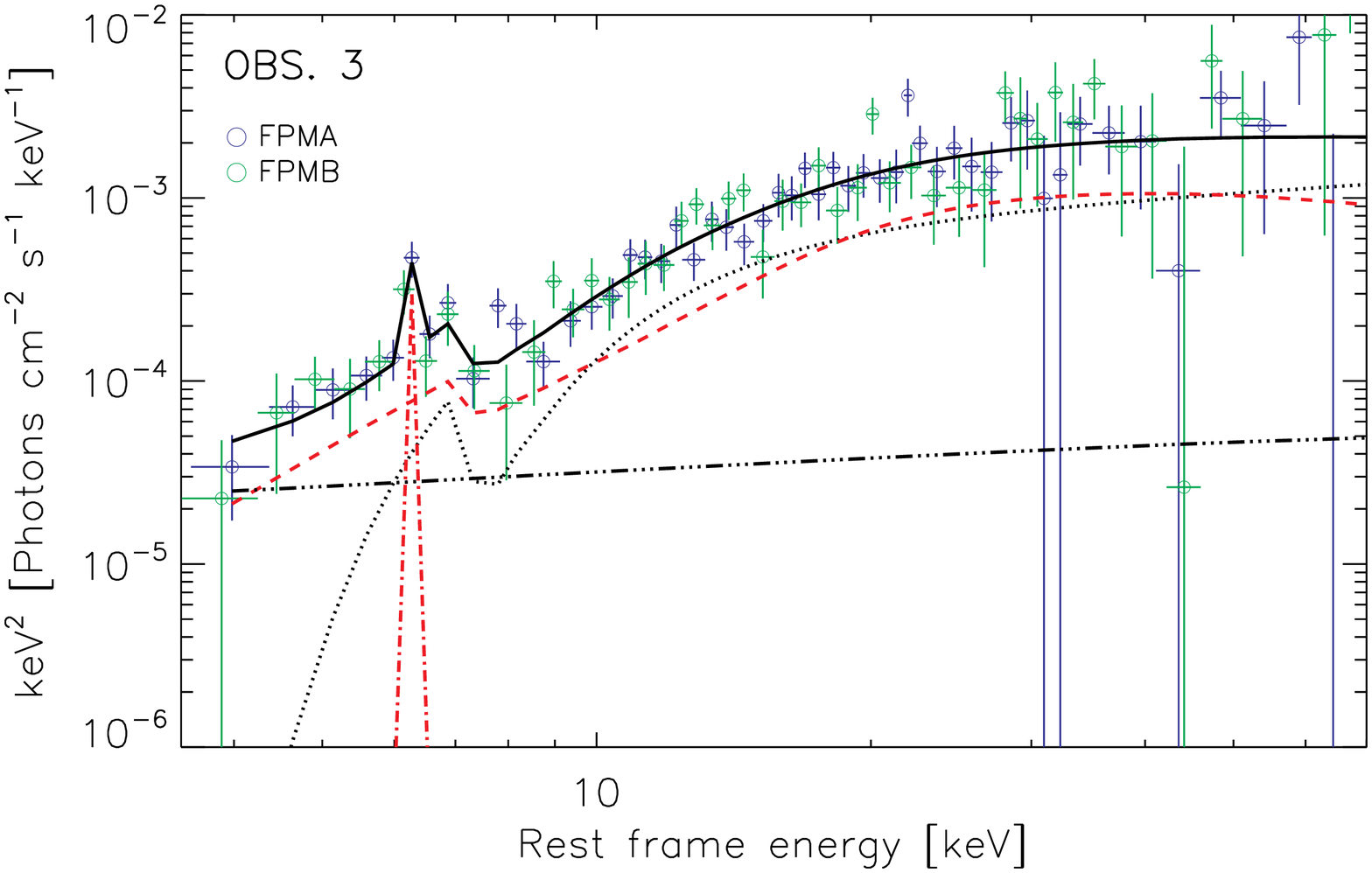}\end{minipage}
\begin{minipage}[!b]{.48\textwidth}
\centering
\includegraphics[width=8cm]{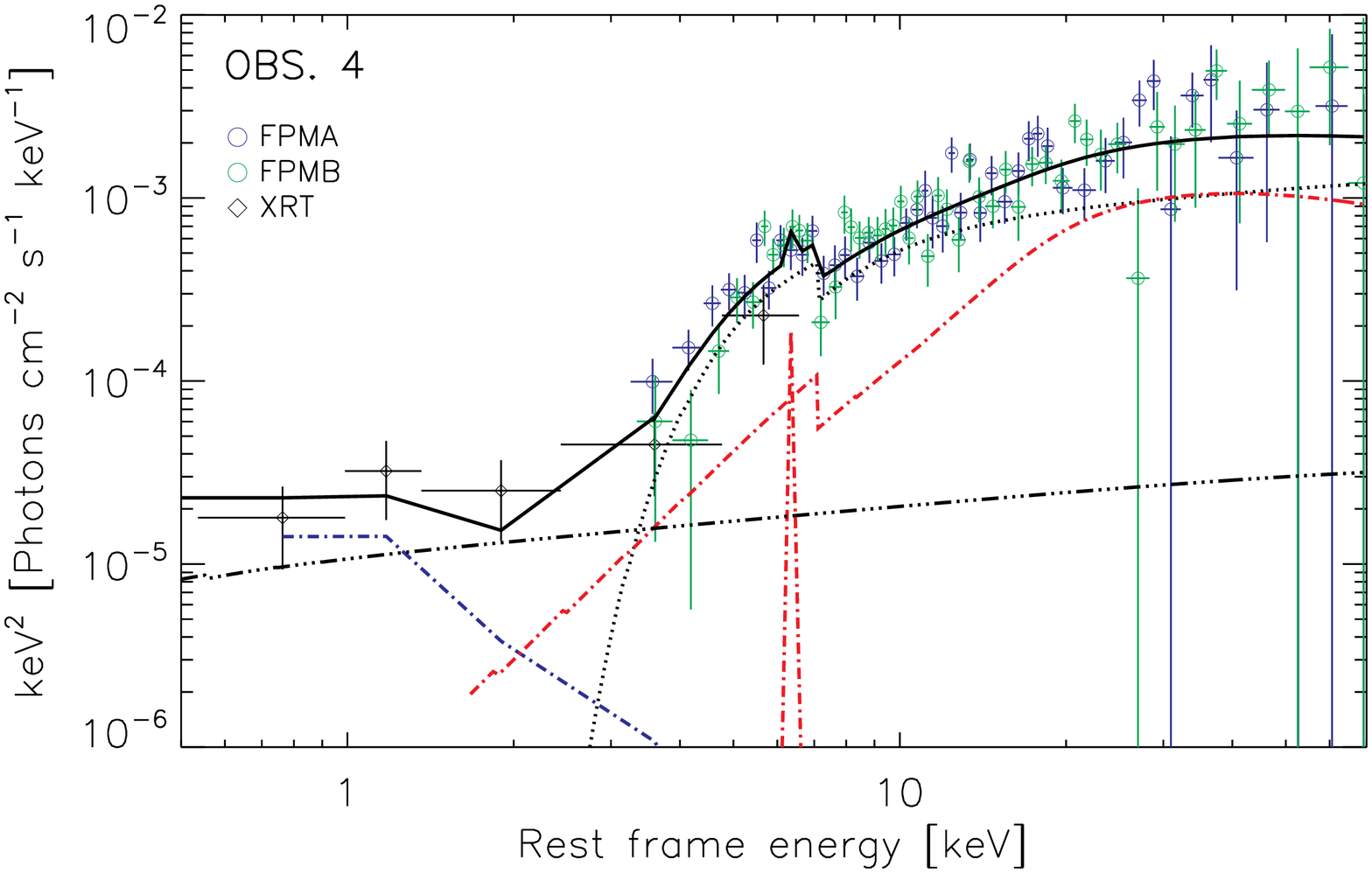}\end{minipage}
\begin{minipage}[!b]{.48\textwidth}
\centering
\includegraphics[width=8cm]{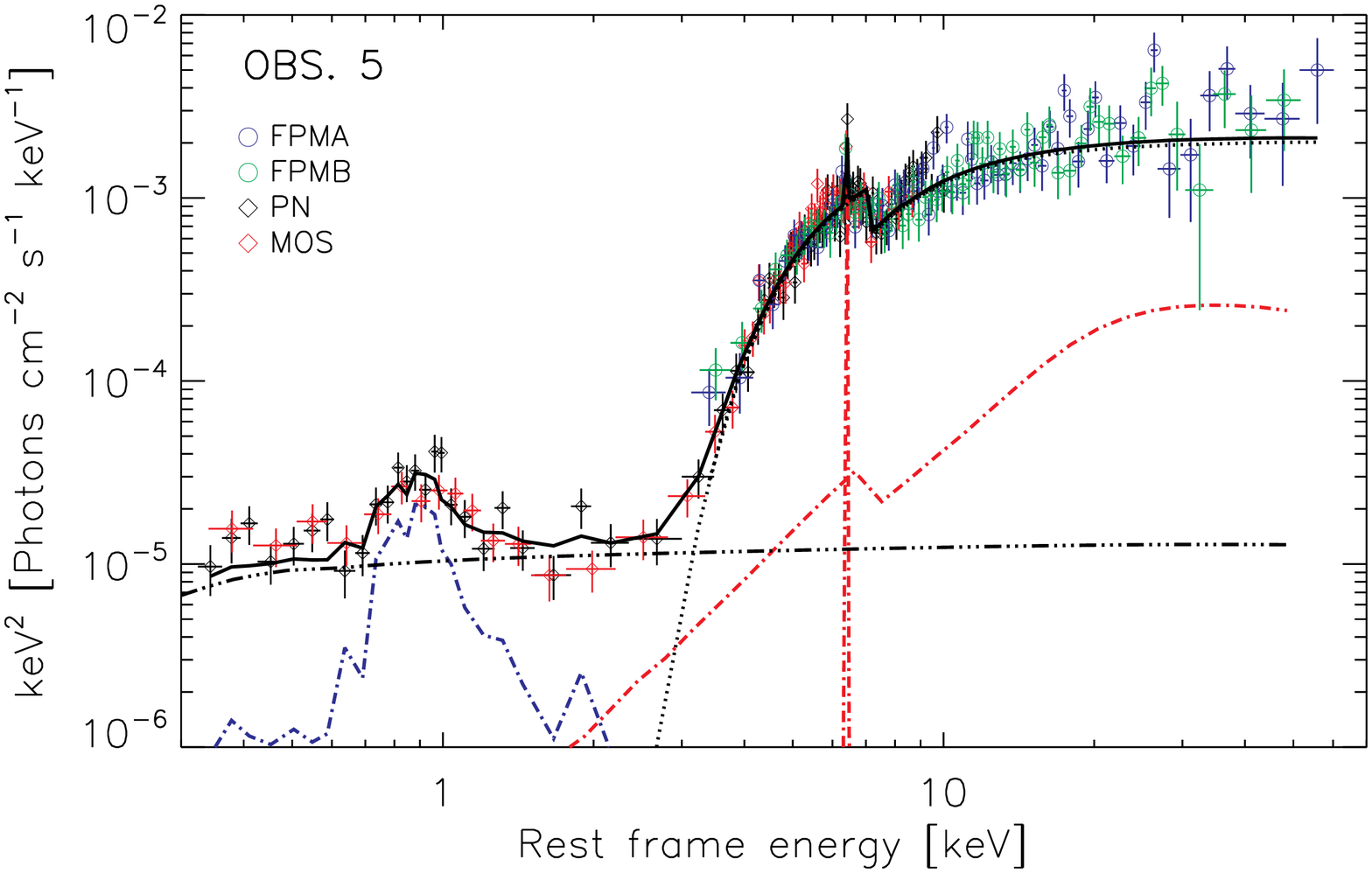}\end{minipage}
\begin{minipage}[!b]{.95\textwidth}
\centering
\includegraphics[width=8cm]{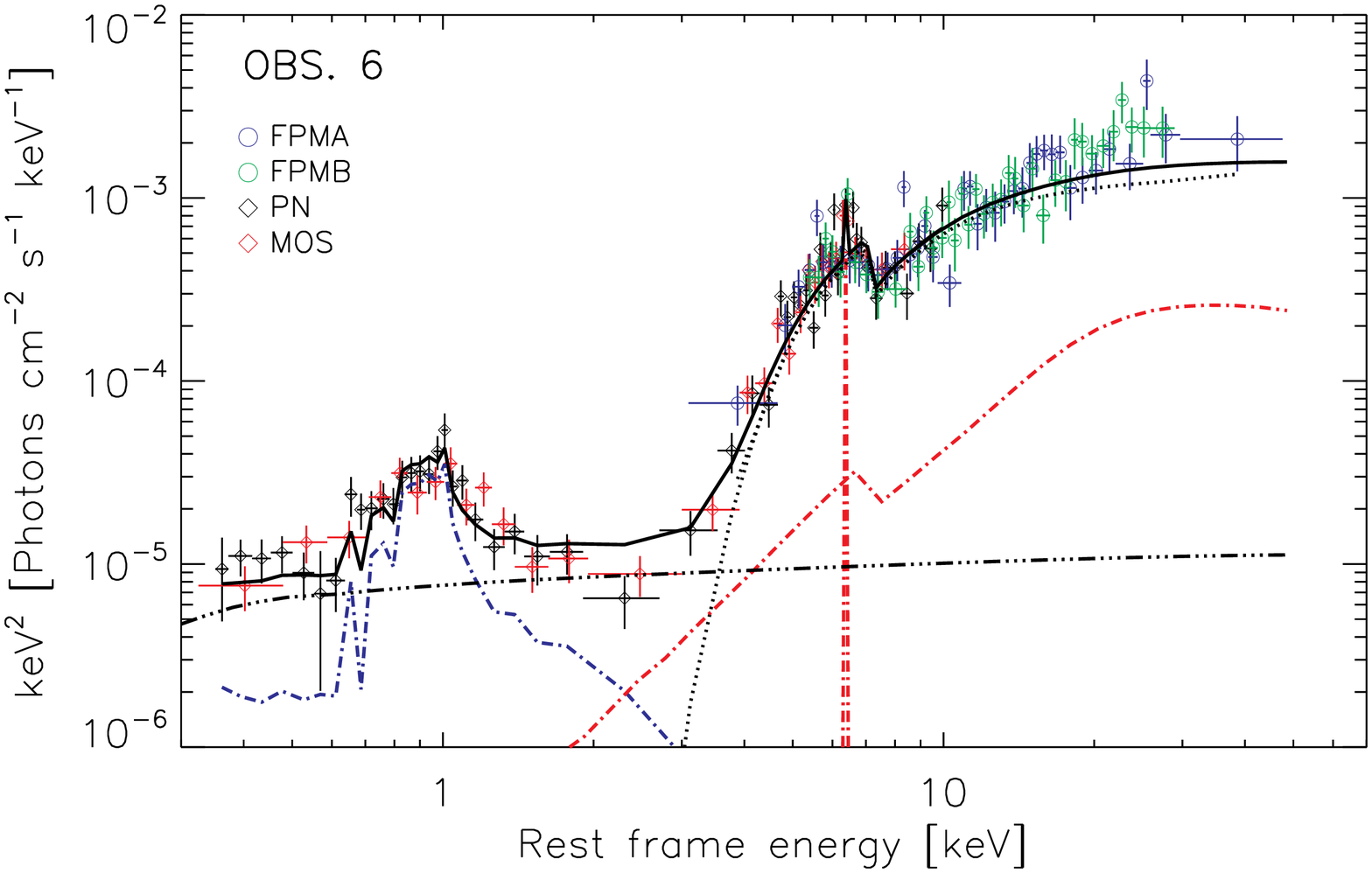}\end{minipage}
\begin{minipage}[!b]{.48\textwidth}
\centering
\includegraphics[width=8cm]{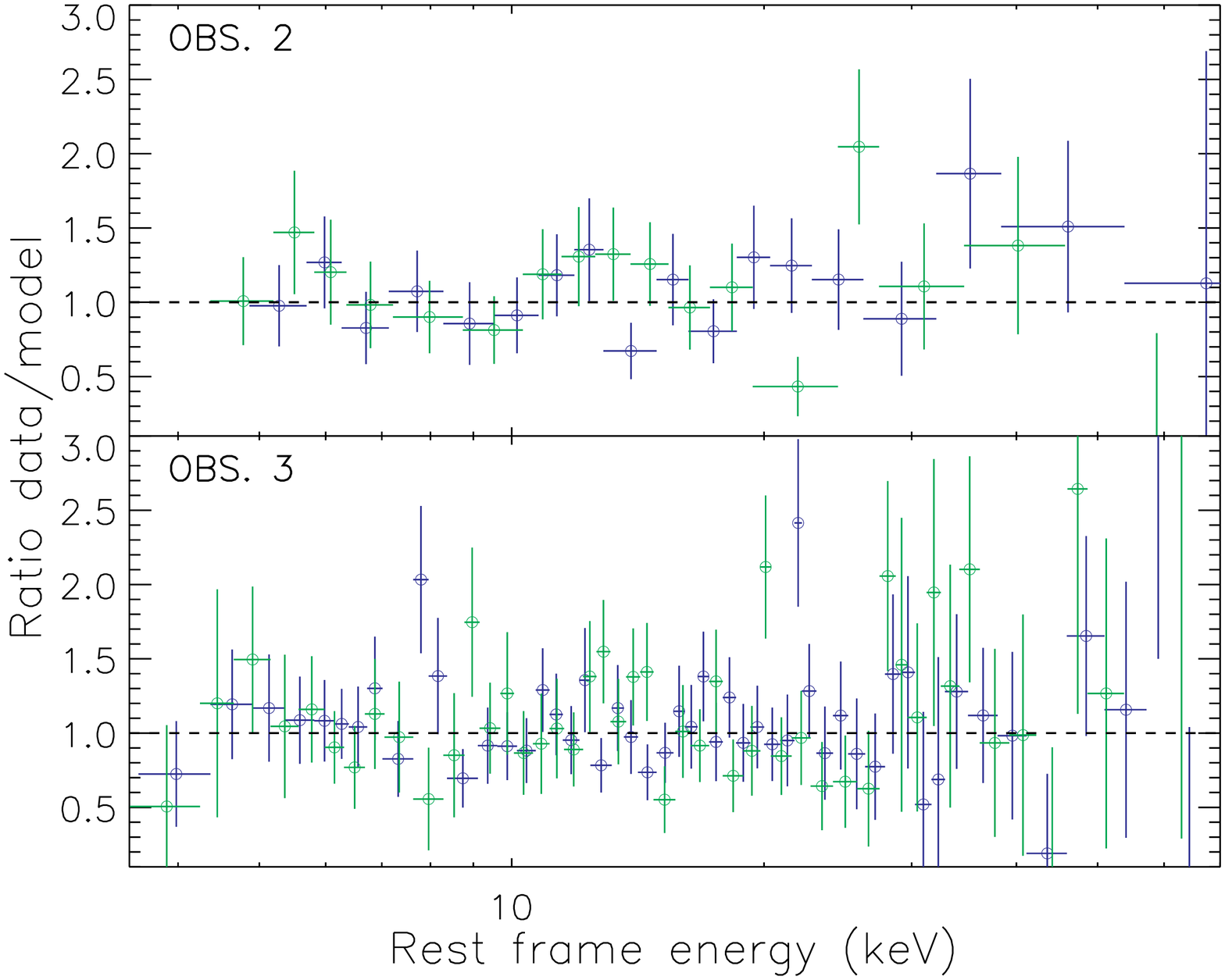}\end{minipage}
\begin{minipage}[!b]{.48\textwidth}
\centering
\includegraphics[width=8cm]{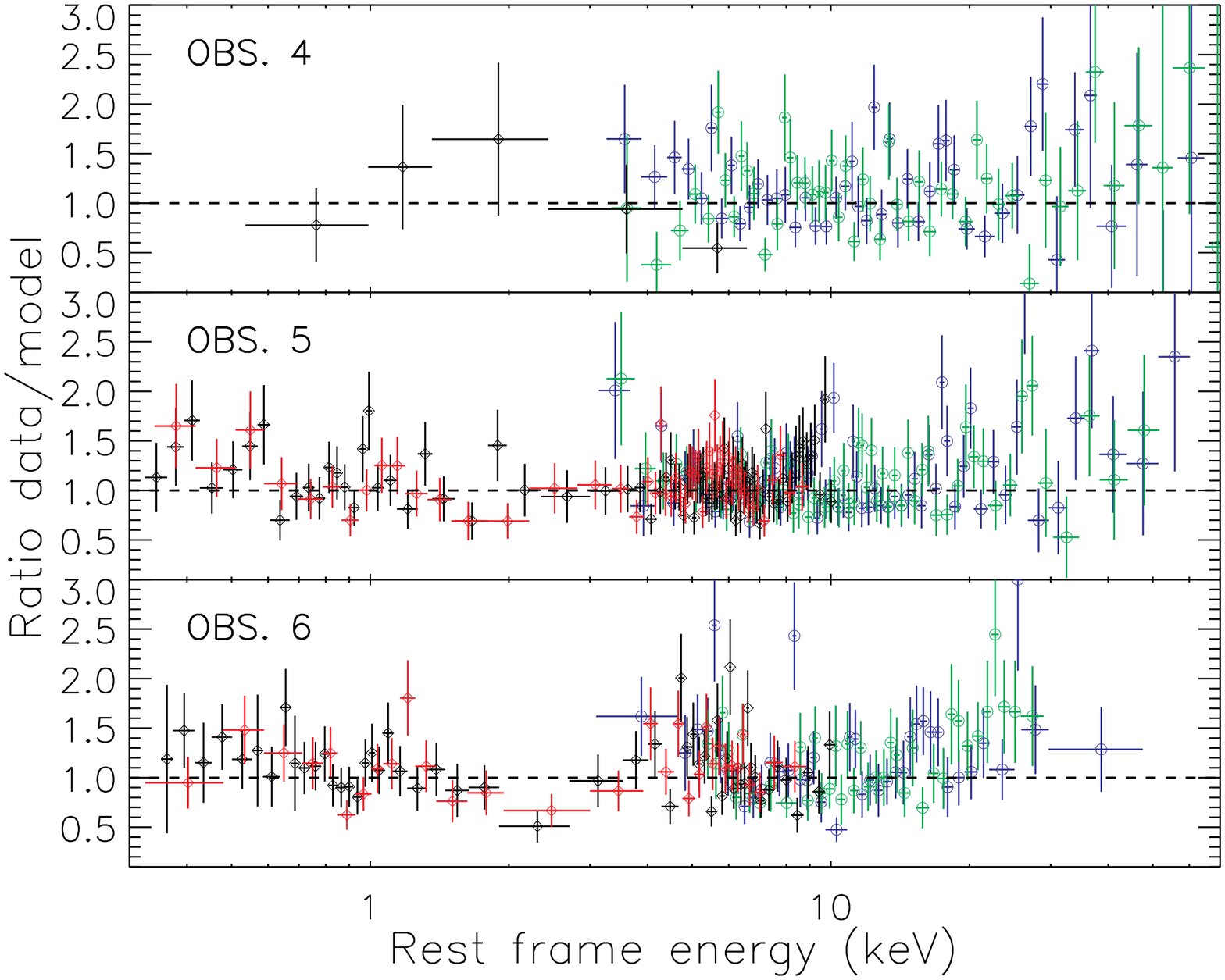}\end{minipage}
 \begin{minipage}[t]{1\textwidth}
  \caption{X-ray spectra of IC\,751 obtained for the five observations discussed here. {\it Swift}/XRT data were rebinned to have a significance of at least 2$\sigma$ per bin only for visual clarity. The black continuous line represents the best fit to the data obtained using the slab model described in Sect.\,\ref{sect:Xrayspecanal1}. The black dotted line represents the absorbed primary cutoff power-law continuum, the red dot-dashed line shows the features arising from reprocessed X-ray emission (Compton hump and Fe K$\alpha$ line), the blue dot-dashed line represents the collisionally ionized plasma emission, while the dot-dot-dashed black line is the scattered component. The bottom panels show the ratio between the data and the model obtained for the five observations (symbols are the same as for the X-ray spectra).}
\label{fig:eeufsobs}
 \end{minipage}
\end{figure*}

\subsection{{\it Swift}}
IC\,751 was observed twice by the X-ray Telescope (XRT, \citealp{Burrows:2005vn}) on board the {\it Swift} observatory \citep{Gehrels:2004dq}: for 2.3\,ks in February 2008 and for 5.8\,ks, two days after the {\it NuSTAR} observation, in May 2013. During the first observation only five counts were detected, so that no detailed spectral analysis could be performed. The data were reduced using the \textsc{xrtpipeline v0.13.0}, which is part of the XRT Data Analysis Software within Heasoft\,v6.16.

The Burst Alert Telescope (BAT) onboard {\it Swift} has been monitoring the sky in the 14--195\,keV band since 2005, and has detected so far more than 800 AGN \citep{Baumgartner:2013uq}, of which 55 CT sources \citep{Ricci:2015fk}. Given the significant $N_{\rm H}$ variability of IC\,751 found by {\it NuSTAR}, we did not use the 70-month stacked {\it Swift}/BAT spectrum for our spectral analysis. The long-term variability inferred by {\it Swift}/BAT will be discussed in Sect.\,\ref{sect:lc}. The 70-month averaged flux of IC\,751 in the 14--195\,keV band is $13.1^{+3.9}_{-3.6}\times 10^{-12}\rm\,erg\,s^{-1}\,cm^{-2}$ (90\% confidence interval).

\section{X-ray spectral analysis -- Slab model}\label{sect:specAnalysis}
We performed X-ray spectral analysis with XSPEC\,v.12.8.2 \citep{Arnaud:1996kx}. To all models we added a photoelectric absorption component (\textsc{tbabs}, \citealp{Wilms:2000vn}) to take into account Galactic absorption in the direction of the source, fixing the value of the column density to $N_{\rm\,H}^{\rm\,G}= 1.2\times 10^{20}\rm\,cm^{-2}$ \citep{Kalberla:2005fk}. In order to use $\chi^2$ statistics, {\it NuSTAR} FPMA/FPMB and {\it XMM-Newton} EPIC PN and MOS spectra were binned to have at least 20 counts per bin. Quoted errors correspond to 90\% confidence level ($\Delta \chi^2=2.7$). Given the low signal-to-noise of the {\it Swift}/XRT spectrum of Obs.\,4 we did not bin the spectrum, and applied Cash statistics (\citealp{Cash:1979fk}, \textsc{cstat} in XSPEC). 

As a first step to shed light on the spectral variability we used a model which considers reprocessed radiation from a slab to reproduce the X-ray spectrum of IC\,751. Since the first two datasets lacked simultaneous coverage below 5\,keV, and the {\it Swift}/XRT observation in Obs.\,4 did not have a very high signal-to-noise ratio, we used the two joint {\it NuSTAR}/{\it XMM-Newton} spectra (obs. 5 and 6, Sect.\,\ref{sect:Xrayspecanal1}) to constrain the fundamental parameters (photon index, normalization of the reflection component, normalization of the Fe K$\alpha$ line), and then used this information to fit the other observations (obs. 2, 3 and 4, Sect.\,\ref{sect:Xrayspecanal2}), in order to constrain the value of the line-of-sight column density, $N_{\rm H}$, and the normalization of the X-ray primary emission. Observation\,1 was not used due to its poor statistics. It was however possible to infer the 2--10\,keV flux of the X-ray source during this observation ($3.1^{+0.4}_{-2.4}\times 10^{-13}\rm\,erg\,s\,cm^{-2}$), which is consistent with that of Observation\,3.

\begin{figure}[t!]
\centering
\includegraphics[width=9cm,angle=0]{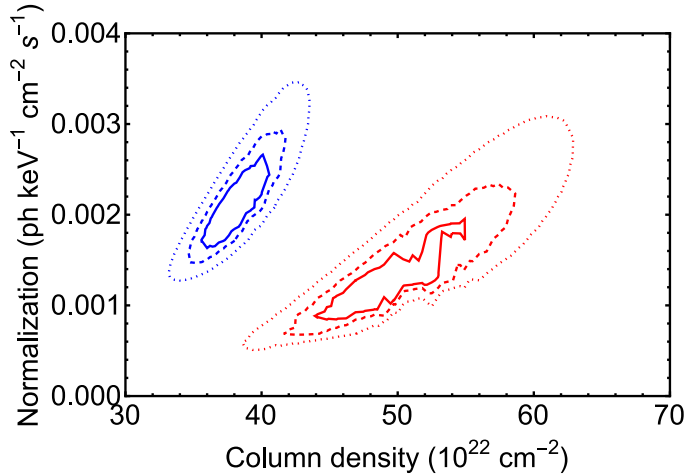}
  \caption{Contour plot of the power-law normalization and the column density for Obs.\,5 (blue) and Obs.\,6 (red). Continuous, dashed and dotted contours represent the 68\%, 90\% and 99\% confidence intervals, respectively.}
\label{fig:contourNHobs56}
\end{figure}

\subsection{Observations 5 and 6}\label{sect:Xrayspecanal1}
The slab model we used to analyze the broad-band X-ray spectrum of IC\,751 includes: i) a power-law with a high-energy cutoff, which represents the primary X-ray emission; ii) photoelectric absorption and Compton scattering, to take into account the line-of-sight obscuration; iii) unabsorbed X-ray reprocessed radiation; iv) a Gaussian line to reproduce the Fe K$\alpha$ emission; v) a cutoff power-law component to reproduce the scattered X-ray emission; vi) a thermal plasma model, to take into account a possible contribution of the host galaxy to the X-ray spectrum below $\sim 2$\,keV. It must be stressed that the emission below 2\,keV, which cannot be accounted for by the scattered component alone, could also be due to the blending of emission lines created by photo-ionisation. However, due to the limited energy resolution of our data, in the following we will use a thermal plasma model. To reproduce photoelectric absorption and Compton scattering we used the \textsc {tbabs} and \textsc{cabs} models, respectively. We took into account the reprocessed X-ray radiation using the \textsc{pexrav} model \citep{Magdziarz:1995pi}, which assumes reflection from a semi-infinite slab. The value of the reflection parameter ($R$) was set to be negative in order to include only the reprocessed component. The values of the photon index ($\Gamma$), cutoff energy ($E_{\rm\,C}$) and the normalization were fixed to the values of the cutoff power-law, while the inclination angle of the observer with respect to the reflecting slab was set to $i=30^{\circ}$. This angle was selected considering the type 2 nature of IC\,751, and assuming that the reflecting material is associated with the molecular torus. The width of the Fe K$\alpha$ line ($\sigma$) was fixed to 40\,eV, a value well below the energy resolution of {\it XMM-Newton} and {\it NuSTAR}, while the energy and normalization of the line were left free to vary. The width of the Fe K$\alpha$ was chosen considering that the bulk of the line is created in material located in the BLR or in the molecular torus (e.g., \citealp{Shu:2010zr}; \citealp{Ricci:2014fk,Ricci:2014fv}). The scattered X-ray emission was taken into account by multiplying an additional unabsorbed cutoff power law by a constant ($f_{\rm\,scatt}$, typically of the order of a few percent), fixing all the parameters to those of the primary X-ray emission. A multiplicative constant to include possible cross-calibration offset between the different instruments was added, and was found to be typically $\lesssim 12\%$ of unity. In XSPEC our model is:
\smallskip

 \textsc{constant$\times$tbabs$_{\rm\,Gal}\times$[ztbabs$\times$cabs$\times$cutoffpl + apec + pexrav + zgauss + $f_{\rm\,scatt}\times$cutoffpl]}.

\smallskip
The results of the spectral fitting are reported in Table\,\ref{tab:resultsFit_XMM_Nustar}. In both Obs.5 and Obs.6 we found the source in a Compton-thin state, with a photon index of $\Gamma \sim 1.9$ and a high-energy cutoff of $E_{\rm\,c}\gtrsim 200$\,keV. The two observations show a change in $N_{\rm H}$ (Fig.\,\ref{fig:contourNHobs56}), with Obs.5 being less obscured than Obs.6 ($\Delta N_{\rm\,H}\sim 10^{23}\rm\,cm^{-2}$,  significant at a $\sim2\sigma$ level). The two observations also show evidence of intrinsic flux variation, with the source being significantly brighter in Obs.5. The reflection component appears to be rather weak, with a value of $R=0.14^{+0.16}_{-0.13}$ in Obs.\,6, while it is less constrained in Obs.\,5 due to the higher flux level. The fraction of scattered flux is found to be $f_{\rm\,scatt}\sim 0.5\%$ of the primary X-ray flux. The X-ray spectra and the ratio between the data and the best-fitting model are shown in Fig.\,\ref{fig:eeufsobs}.

\subsection{Observation 2, 3 and 4}\label{sect:Xrayspecanal2}

To study Obs.\,2, 3 and 4 we set all the parameters, with the exception of the normalization of the primary X-ray emission and the column density, to the values obtained from the study of Obs.\,6. The normalization and $N_{\rm\,H}$ were allowed to vary within the uncertainties of the values obtained by fitting the X-ray spectrum of Obs.\,6. Observation 6 was chosen because the X-ray source was caught in a low-flux state, which allows better constraints on the normalization of the Fe K$\alpha$ line and of the reprocessed X-ray emission.

Fitting Obs.2 with the approach described above we obtained a chi-squared of 49.5 for 28 DOF, and a clear excess below 6\,keV. This excess can be removed by allowing the obscuring material to partially cover the X-ray source, so that some of the primary X-ray flux is able to leak out unabsorbed; i.e., by leaving $f_{\rm\,scatt}$ free to vary. This model yields a good value of chi-squared ($\chi^2$/DOF=31.3/27), and a fraction of unabsorbed flux of $\sim 17\%$. Assuming that the scattered fraction is $\sim 0.5\%$, as found by the spectral analysis of Obs.\,5 and 6, this would imply that the absorber is covering $\sim 83\%$ of the X-ray source in the line-of-sight. The column density is significantly larger than in Obs.\,5 and 6, with the obscurer consistent with being CT [$\log (N_{\rm\,H}/\rm cm^{-2})= 24.3$], while the source is in a high-flux state during this observation (Table\,\ref{tab:resultsFit_XMM_Nustar}).

The X-ray spectrum of Obs.3 shows that the X-ray source was obscured by CT material also $\sim3$\,months later [$\log (N_{\rm\,H}/\rm cm^{-2})\sim24.13$]. As can be seen in Fig.\,\ref{fig:eeufsobs}, the source was not in a high-flux state anymore, and the spectrum shows a prominent Fe K$\alpha$ line (EW$\sim 500$\,eV). Leaving the value of $f_{\rm\,scatt}$ free to vary does not improve significantly the chi-squared ($\Delta \chi^2\sim 1$ for 1 less DOF). The Fe K$\alpha$ EW is higher in Obs.\,3 than in Obs.\,2 due to the higher flux level of the X-ray source during Obs.\,2.

The broad-band {\it Swift}/XRT--{\it NuSTAR} spectrum of Obs.\,4 shows that IC\,751 was in a Compton-thin state at the end of May 2013, while the X-ray source was significantly dimmer with respect to the previous two observations, with its intrinsic flux level comparable to that observed during Obs.\,6.

\section{X-ray spectral analysis -- Torus model}\label{sect:physical modelling}

To investigate further the structure of the absorbers, and to disentangle the torus absorption from that caused by clouds in the line-of-sight, assuming an homogeneous torus, we used the \textsc{MYTorus} model\footnote{http://www.mytorus.com/} \citep{Murphy:2009ly}. The \textsc{MYTorus} model considers absorbed and reprocessed X-ray emission from a smooth torus with a half-opening angle $\theta_{\mathrm{OA}}$ of $60^{\circ}$, and can be used for spectral fitting as a combination of three additive and exponential table models. These tabulated models include the zeroth-order continuum\footnote{ This component takes into account both Compton scattering and photoelectric absorption.} (\textsc{mytorusZ}), the scattered continuum (\textsc{mytorusS}) and a component which contains the fluorescent emission lines (\textsc{mytorusL}). 

\begin{figure}[t!]
\centering
\centering
\includegraphics[height=9cm,angle=270]{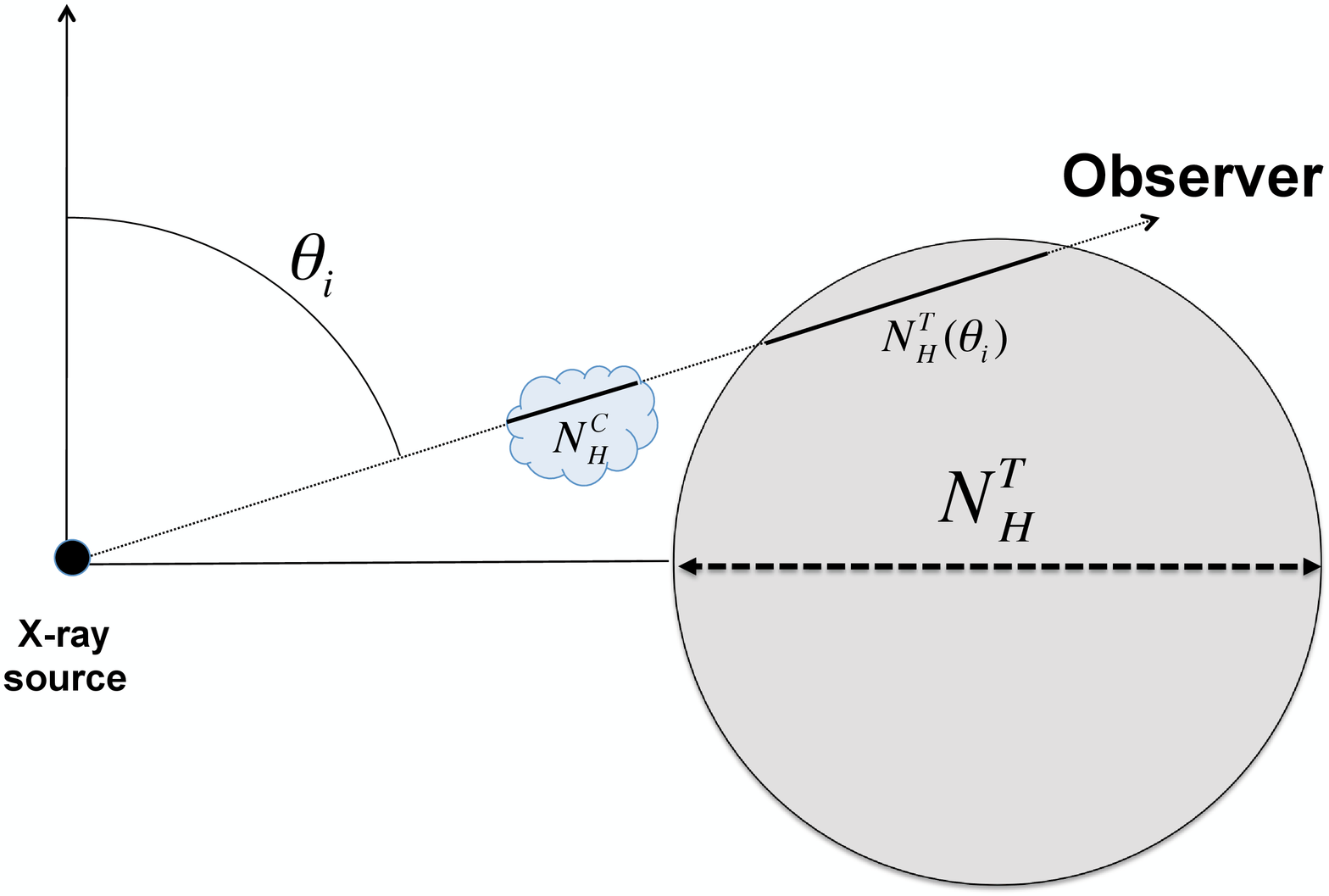}
  \caption{Geometry assumed for the torus model reported in Sect.\,\ref{sect:standmyt}. The parameters $\theta_{\rm\,i}$, $N_{\rm\,H}^{\rm\,C}$, $N_{\rm\,H}^{\rm\,T}(\theta_{\rm\,i})$ and $N_{\rm\,H}^{\rm\,T}$ are the inclination angle, the column density of the cloud (variable), the column density of the torus at an angle $\theta_{\rm\,i}$ (non-variable), and the equatorial (i.e. maximum) column density of the torus, respectively. }
\label{fig:physicalGeom}
\end{figure}

\subsection{Standard MYTorus}\label{sect:standmyt}
The analysis we carried out using the slab model (Sect.\,\ref{sect:specAnalysis}) showed that $N_{\rm H}$ is highly variable, so that it cannot be associated with a smooth absorber alone. This is also confirmed by the fact that applying the smooth \textsc{MYTorus} model to all the X-ray spectra available, setting the values to be the same for the different observations, results in a chi-squared of $\chi^{2}=2628.2$ for 707 DOF. Also considering different normalisations of the direct and scattered component or different values of the column density of the scattering and absorbing material fails to reproduce the X-ray spectrum, resulting in values of the reduced chi-squared of $\chi^{2}_{\nu}>2$. We therefore used an alternative approach to take into account variable absorption by combining the non-varying torus absorption [\textsc{mytorusZ(Tor)}] with what we define as the {\it cloud absorption}. The geometry of the absorber we assume is shown in Fig.\,\ref{fig:physicalGeom}. We adopted a model which includes the three components of \textsc{MYTorus} plus a collisionally ionized plasma, and power law to reproduce the scattered component, similar to what was done in Sect.\,\ref{sect:specAnalysis}. In order to take into account the variable absorber we used an additional obscuring multiplicative component [\textsc{mytorusZ(Cloud, $N_{\rm\,H}^{\rm\,C}$)}], where $N_{\rm\,H}^{\rm\,C}$ is the column density of the cloud. In XSPEC the syntax of our model is: 
\smallskip

\textsc{constant$\times$tbabs$_{\rm\,Gal}\times$[mytorusZ(Cloud) $\times$mytorusZ(Tor)$\times$ zpowerlaw + mytorusS + apec + gsmooth(mytorusL) + $f_{\rm\,scatt}\times$zpowerlaw]}.

\smallskip

The free parameters of \textsc{MYTorus} are the photon index $\Gamma$, the equatorial column density of the torus $N_{\rm\,H}^{\mathrm{\,T}}$ and the inclination angle of the observer $\theta_{\rm\,i}$. We convolved the fluorescent emission lines of \textsc{MYTorus} using a Gaussian function (\textsc{gsmooth} in XSPEC) to take into account the expected velocity broadening. We fixed the width of the lines to a full width half maximum of $\mathrm{FWHM}=2000\rm\,km\,s^{-1}$, consistent with the average value obtained for 36 AGN at $z<0.3$ by the {\it Chandra}/HEG study of \cite{Shu:2010zr}. The model was also multiplied by a constant to take into account cross-calibration between the different spectra.

\begin{table*}
\begin{center}
\caption[]{X-ray spectral analysis -- torus model}
\label{tab:resultsFit_physical}
\begin{tabular}{lcccccccc}
\noalign{\smallskip}
\hline \hline \noalign{\smallskip}
\multicolumn{9}{c}{{\normalsize MYTorus}} \\
\hline \noalign{\smallskip}
\multicolumn{1}{l}{\,\,\,\,\,\,\,\,\,\,\,\,(1)} & (2) &  (3) &  (4) &  (5) &  (6) &  (7) &  (8) &  (9)   \\
\noalign{\smallskip}
Observation & $N_{\rm\,H}^{\rm\,C}$ & $n_{\rm\,po}$ & $\Gamma$ & $N_{\rm\,H}^{\rm\,T}$  & $\theta_{\rm\,i}$& $n_{\rm\,refl}$  & kT  & $f_{\rm\,scatt}$  \\
\noalign{\smallskip}
 & [{\tiny $10^{22}\rm\,cm^{-2}$}] & [{\tiny $10^{-3} \rm\,ph\,keV^{-1}\,cm^{-2}\,s^{-1}$}] & &  [{\tiny $10^{24}\rm\,cm^{-2}$}]  & [{\tiny deg}] & [{\tiny $10^{-3}\rm\,ph\,keV^{-1}\,cm^{-2}\,s^{-1}$}] & [{\tiny keV}] & [{\tiny \%}] \\
\noalign{\smallskip}
\noalign{\smallskip}
\hline \noalign{\smallskip}
\textbf{\small 2} [2012.82]	 & $150^{+19}_{-15}$  &$3.83^{+0.43}_{-0.42}$  	&$1.88^{+0.01}_{-0.04}$ &$4.76^{+0.09}_{-0.27}$ &$60.3^{+0.2}_{-0.2}$  	& $1.14^{+0.15}_{-0.11}$ &$0.93^{+0.04}_{-0.04}$ & $4.2^{+0.8}_{-0.8}$\\
\noalign{\smallskip}
\textbf{\small 3} [2013.10]	 & $111^{+8}_{-7}$  & $2.10^{+0.18}_{-0.17}$	 &// &// &//  	&// &// & $1.6^{+0.5}_{-0.5}$\\
\noalign{\smallskip}
\textbf{\small 4} [2013.39]	 &$5.1^{+3.0}_{-3.1}$   & $0.94^{+0.06}_{-0.06}$	 & // &// & // 	&//  & // &$3.2^{+1.5}_{-1.2}$ \\
\noalign{\smallskip}
\textbf{\small 5} [2014.91]	 &  $0.58^{+0.08}_{-0.05}$  & $1.83^{+0.07}_{-0.05}$ 	 &//  &//  &//  	& //  &//  & $0.4^{+0.1}_{-0.1}$ \\
\noalign{\smallskip}
\textbf{\small 6} [2014.92]	 & $0.49^{+0.10}_{-0.05}$  & $0.75^{+0.04}_{-0.04}$ 	 &// &// &// 	& // &// &$0.8^{+0.1}_{-0.1}$ \\
\noalign{\smallskip}
\hline
\noalign{\smallskip}
\noalign{\smallskip}
\multicolumn{9}{c}{{\normalsize MYTorus -- decoupled model}} \\
\hline \noalign{\smallskip}
Observation & $N_{\rm\,H}^{\rm\,C}$ & $n_{\rm\,po}$ & $\Gamma$ & $N_{\rm\,H}^{\rm\,T}(Z)$  & $N_{\rm\,H}^{\rm\,T}(S,L)$ & $n_{\rm\,refl}$  & kT  & $f_{\rm\,scatt}$  \\
\noalign{\smallskip}
 & [{\tiny $10^{22}\rm\,cm^{-2}$}] & [{\tiny $10^{-3} \rm\,ph\,keV^{-1}\,cm^{-2}\,s^{-1}$}] & &  [{\tiny $10^{24}\rm\,cm^{-2}$}]  &  [{\tiny $10^{24}\rm\,cm^{-2}$}]& [{\tiny $10^{-3}\rm\,ph\,keV^{-1}\,cm^{-2}\,s^{-1}$}] & [{\tiny keV}] & [{\tiny \%}] \\
\noalign{\smallskip}
\noalign{\smallskip}
\hline \noalign{\smallskip}
\textbf{\small 2} [2012.82]	 & $147^{+82}_{-48}$  &$4.71^{+6.29}_{-2.45}$  	&$1.98^{+0.08}_{-0.07}$ &$0.37^{+0.02}_{-0.02}$ &$6.0^{+3.5}_{-2.2}$  	& $1.22^{+0.50}_{-0.36}$ &$0.93^{+0.05}_{-0.11}$ & $4.1^{+3.4}_{-2.0}$\\
\noalign{\smallskip}
\textbf{\small 3} [2013.10]	 & $105^{+32}_{-26}$  & $2.43^{+1.60}_{-0.90}$	 &// &// &//  	&// &// & $1.6^{+0.9}_{-0.7}$\\
\noalign{\smallskip}
\textbf{\small 4} [2013.39]	 &$\leq 18$   & $1.20^{+0.45}_{-0.30}$	 & // &// & // 	&//  & // &$2.7^{+3.4}_{-1.8}$ \\
\noalign{\smallskip}
\textbf{\small 5} [2014.91]	 &  $0.64^{+0.17}_{-0.07}$  & $2.08^{+0.05}_{-0.03}$ 	 &//  &//  &//  	& //  &//  & $0.4^{+0.1}_{-0.1}$ \\
\noalign{\smallskip}
\textbf{\small 6} [2014.92]	 & $11.1^{+5.0}_{-4.9}$  & $1.28^{+0.38}_{-0.27}$ 	 &// &// &// 	& // &// &$0.6^{+0.2}_{-0.3}$ \\

\noalign{\smallskip}
\hline
\noalign{\smallskip}

\end{tabular}
\tablecomments{Model parameters obtained by fitting simultaneously the 15 X-ray spectra of IC\,751 (divided in 5 epochs).  The table reports (1) the observation number, (2) the cloud column density, (3) the normalization of the primary power-law emission, (4) the photon index of the primary X-ray emission, (5) the equatorial column density of the torus, (6) the inclination angle of the observer (see Fig.\,\ref{fig:physicalGeom}), (7) the normalization of the reflection component, (8) the temperature of the collisionally ionized plasma, (9) the fraction of scattered unabsorbed emission. The upper part of the table refers to the \textsc{MYTorus} model in his original formulation (with the addition of a cloud of neutral material, see Sect.\,\ref{sect:standmyt}), while the lower part reports the results obtained by using \textsc{MYTorus} in the decoupled mode (Sect.\,\ref{sect:mytdec}). In the lower part column (5) is the torus column density of the \textsc{mytorusZ} component for $\theta_{\rm\,i}(Z)=90^{\circ}$, and (6) is the column density of the \textsc{mytorusS} and \textsc{mytorusL} components, assuming $\theta_{\rm\,i}(S,L)=0^{\circ}$ and $\theta_{\rm\,i}(S,L)=90^{\circ}$.}
\end{center}
\end{table*}

\begin{figure}[t]
\centering
\includegraphics[width=8.5cm]{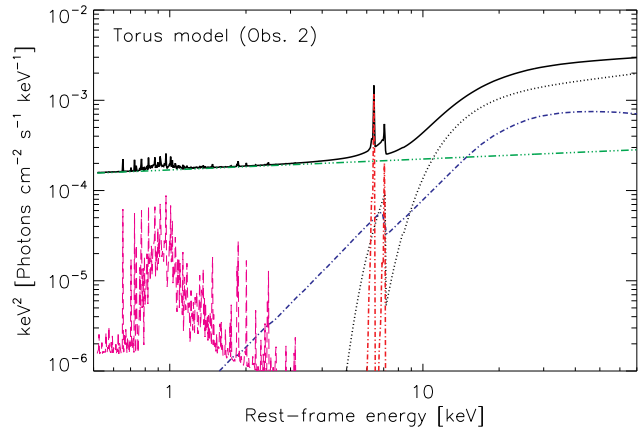}
\caption{Torus spectral model used for the analysis of IC\,751. The black continuous line represents the total flux, while the components shown are: the absorbed X-ray power-law (\textsc{zpow$\times$ztbabs$\times$cabs$\times$mytorusZ}, dotted black line), the scattered component from the torus (\textsc{mytorusS}, dot-dashed blue line), the fluorescent emission lines (\textsc{mytorusL}, dot-dashed red line), the thermal component (magenta dashed line), and the scattered emission (green dot-dot-dashed line). The values are set to those obtained during Obs.\,2, for more details see Sect.\,\ref{sect:standmyt} and Table\,\ref{tab:resultsFit_physical}.}
\label{fig:mytmodel}
\end{figure}

We fitted simultaneously the five sets of observations discussed above. The values of $\theta_{\rm\,i}$, $N_{\rm\,H}^{\rm\,T}$, $\Gamma$, and the normalization of the scattered component ($n_{\rm\,refl}$, which includes the Compton hump) were left free to vary, and tied to be constant for all observations. The normalization of the fluorescent-lines was fixed to $n_{\rm\,refl}$. The normalization of the primary power-law component ($n_{\rm\,po}$), the value of $N_{\rm\,H}^{\rm\,C}$ and that of $f_{\rm\,scatt}$ were left free to have independent values for different observations.

The results obtained by this torus model are reported in the upper part of Table\,\ref{tab:resultsFit_physical}. The fit yields a chi-squared of $\chi^{2}=774.8$ (for 688 DOF), and the primary X-ray emission has a value of the photon index consistent with that found using the slab model. As with the slab model, we find a clear variation of the line-of-sight column density, with the clouds having values of the column density spanning between $1.5\times 10^{24}$ and $5\times 10^{21}\rm\,cm^{-2}$. This approach also confirms a significant change in the value of $f_{\rm\,scatt}$ between the five observations, which varies between $\sim 4\%$ and $\sim 0.4\%$. These variations in $f_{\rm\,scatt}$ are interpreted as being due to a partially covering absorber in the line of sight. Contrary to what we obtained using the slab model, applying \textsc{MYTorus} we do not find a significant variation of the line-of-sight column density between Obs.\,5 and Obs.\,6. The model used, with the parameters set to those obtained for Obs.\,2, is shown in Fig.\,\ref{fig:mytmodel}.

We found an equatorial column density of $N_{\rm\,H}^{\rm\,T}\simeq 4.8\times 10^{24}\rm\,cm^{-2}$ and an inclination angle of the observer of $\theta_{\rm\,i}\simeq 60.3^{\circ}$, close to grazing incidence. The line-of-sight column density in the toroidal geometry assumed by \textsc{MYTorus} can be obtained by 
\begin{equation}\label{eqNH}
N_{\rm\,H}^{\rm\,T}(\theta_{\rm\,i})=N_{\rm\,H}^{\rm\,T} (1-4\cos^{2}\theta_{\rm\,i})^{\frac{1}{2}},
\end{equation}
which implies that for IC\,751 $N_{\rm\,H}^{\rm\,T}(\theta_{\rm\,i})\simeq 6.4\times 10^{23}\rm\,cm^{-2}$, a value larger than the lowest column density inferred by adopting the slab model. However, given the dependence of the column density on the inclination angle, and the problems associated with the toroidal geometry for inclination angles close to the edges (see discussion in \citealp{Yaqoob:2012uq}), the uncertainty associated with this value is large.

\subsection{Decoupled MYTorus}\label{sect:mytdec}
To test further the structure of the absorber, we applied \textsc{MYTorus} in the {\it decoupled} mode \citep{Yaqoob:2012uq}. This was done by: i) separating the column density of the absorbing [$N_{\rm\,H}^{\rm\,T}(Z)$] and reprocessing [$N_{\rm\,H}^{\rm\,T}(S,L)$] material, leaving both of them free to vary; ii) fixing the inclination angle of \textsc{mytorusL} and \textsc{mytorusS} to $\theta_{\rm\,i}(S,L)=0^{\circ}$, and that of \textsc{mytorusZ} to $\theta_{\rm\,i}(Z)=90^{\circ}$; iii) adding a second scattered component with $\theta_{\rm\,i}(S,L)=90^{\circ}$ and the same column density and normalization of the one with $\theta_{\rm\,i}(S,L)=0^{\circ}$; iii) leaving the normalizations of the two components ($n_{\rm\,po}$ and $n_{\rm\,refl}$) free, as was done in Sect.\,\ref{sect:standmyt}. The model we used assumes a geometry which consists of two absorbers, one varying and one constant with time, plus reprocessing material with a different value of the column density.

The results obtained are reported in the lower part of Table\,\ref{tab:resultsFit_physical}. The decoupled \textsc{MYTorus} model yields a better chi-squared ($\chi^{2}=749.2$) than the non-decoupled one, for the same number of DOF. The values of $f_{\rm\,scatt}$ are consistent with those found by applying the standard \textsc{MYTorus} model, while power-law continuum is slightly steeper. With this model we find a significant variation of $N_{\rm\,H}^{\rm\,C}$ between Obs.5 and 6, similarly to what we found using \textsc{pexrav}.

The column density of the reprocessing material is found to be $N_{\rm\,H}^{\rm\,T}(S,L)=(6.0^{+3.5}_{-2.2})\times 10^{24}\rm\,cm^{-2}$, while the line of sight non-variable absorber [$N_{\rm\,H}^{\rm\,T}(Z)$] has a lower value ($\sim 3.7\times 10^{23}\rm\,cm^{-2}$) than that obtained considering a homogeneous torus (Sect.\,\ref{sect:standmyt}). The value of $N_{\rm\,H}^{\rm\,T}(Z)$ is consistent with the lowest value of $N_{\rm\,H}$ obtained by applying the slab model.

\section{Flux Variability and time-resolved spectroscopy}\label{sect:lc}

\begin{figure}[t!]
\centering
\includegraphics[width=9cm]{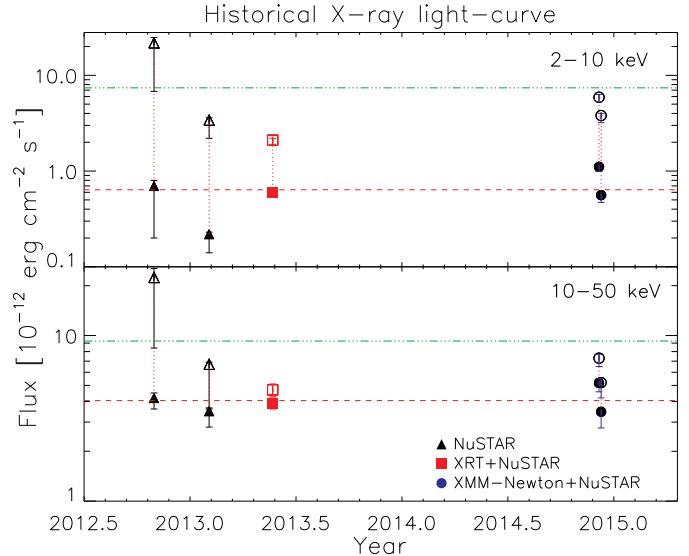}
%
  \caption{Historical X-ray variability of IC\,751 in the 2--10 ({\it top panel}) and 10--50\,keV ({\it bottom panel}) band. The filled and empty points represent the observed and intrinsic (i.e. corrected for absorption) fluxes, respectively. The red dashed line and the dot-dot-dashed green line represent the average observed and intrinsic fluxes, respectively.}
\label{fig:longtermlc}
\end{figure}

X-ray observations have shown that, besides a highly variable line-of-sight column density, IC\,751 also presents significant flux variability of the primary X-ray source on days-to-months timescales both in the soft and hard X-ray bands (see Table\,\ref{tab:resultsFit_XMM_Nustar} and \ref{tab:resultsFit_physical}).  As illustrated in Fig.\,\ref{fig:longtermlc} (filled points), the observed flux of  IC\,751 varies by a factor of four in the 2--10\,keV band, and by a factor of $1.6$ in the 10--50\,keV band. When considering intrinsic fluxes (i.e. absorption corrected, empty points in Fig.\,\ref{fig:longtermlc}), the amplitude of the variability is larger: the X-ray source varies by a factor of $\sim 5$ both in the 2--10 and 10--50\,keV bands. The average observed fluxes in the 2--10\,keV and 10--50\,keV bands are $6.4\times 10^{-13}\rm\,erg\,cm^{-2}\,s^{-1}$ and $4.0\times 10^{-12}\rm\,erg\,cm^{-2}\,s^{-1}$, respectively. The intrinsic average nuclear fluxes in the two bands are $7.4\times 10^{-12}\rm\,erg\,cm^{-2}\,s^{-1}$ (2--10\,keV) and $9.3\times 10^{-12}\rm\,erg\,cm^{-2}\,s^{-1}$ (10--50\,keV), which correspond to k-corrected average luminosities of $\log (L_{2-10}/\rm\,erg\,s^{-1})=43.22$ and $\log (L_{10-50}/\rm\,erg\,s^{-1})=43.32$. The 12$\mu$m rest-frame luminosity of IC\,751 could be estimated using {\it WISE}, by linearly interpolating the fluxes in the $W3$ (11.56\,$\mu$m) and $W4$ (22.09\,$\mu$m) bands. The mid-IR flux of IC\,751 is dominated by the AGN, since $W1-W2>0.8$ \citep{Stern:2012fk}. We found that the X-ray luminosity is in agreement with the 12$\mu$m luminosity [$\log (L_{12\mu\rm m}/\rm\,erg\,s^{-1})=43.70$], as expected from the well known mid-IR/X-ray correlation (e.g., \citealp{Gandhi:2009zr}, \citealp{Stern:2015ve}, \citealp{Asmus:2015ly}).

In order to improve our constraints on the absorbing material, and to study its evolution on shorter timescales, we analysed the {\it XMM-Newton} EPIC/PN and {\it NuSTAR} light-curves of IC\,751 in different energy bands. We extracted {\it XMM-Newton} EPIC/PN light curves in the 0.3--10, 0.3--2 and 2--10\,keV bands, with bins of 1\,ks. {\it NuSTAR} FPMA light curves were extracted in the 3--79, 3--20 and 20--60\,keV bands with bins of 6\,ks. We also analysed the variability of the hardness ratio, defined as $HR=H-S/H+S$, where $H$ and $S$ are the fluxes in the soft (0.3--2 and 3--20\,keV) and hard (2--10 and 20--60\,keV) bands, respectively. 

For all the observations we performed a $\chi^2$ test in order to assess the variability of the hardness ratio and the flux in the three different energy bands. We considered the flux or the hardness ratio to be variable if the minimum confidence level was $p \leq 1\%$. To constrain the amplitude of the variability we used the rms variability amplitude ($F_{\rm\,var}$, see Eq. 10 and B2 of \citealp{Vaughan:2003uq}). {\it NuSTAR} light-curves of Obs.\,2 show significant variability both in the broad and in the soft band, while the other {\it NuSTAR} observations do not show any sign of short-term variability. The {\it XMM-Newton} light-curve of Obs.\,5 shows significant flux variation in the hard band, while the flux is consistent with being constant during Obs.\,6. The largest value of the rms variability amplitude is found for the {\it NuSTAR} 3--20\,keV light-curve of Obs.\,2 ($F_{\rm\,var}= 34\pm7\%$). The hardness ratio is significantly variable only for the {\it NuSTAR} light-curve of Obs.\,2 ($p\simeq 0.5\%$, $F_{\rm\,var}=59\pm16\%$), which shows a hardening of the spectrum in the last $\sim 6$\,ks of the observation. 

The 104-months 14-195 {\it Swift}/BAT light-curve does not show evidence of significant long-term variability ($p\sim 6\%$) on a time-scale of 25\,Ms.

We also carried out time-resolved spectral analysis of the longest {\it NuSTAR} observation (Obs.\,3), by splitting it in five time intervals with similar length. We applied the slab model described in Sect.\,\ref{sect:specAnalysis}, leaving the normalization of the primary X-ray emission and the line-of-sight column density free to vary, and found that the parameters obtained are consistent within their 90\% confidence interval between all observations. Leaving the value of $f_{\rm\,scatt}$ free to vary improves significantly the fit only for the second ($\Delta \chi^2\simeq 8$) and third ($\Delta \chi^2\simeq 7$) segment, and results in values consistent with those obtained for the other segments and for the whole observation.

\medskip
\medskip
\medskip
\medskip

\section{Discussion}

\begin{figure}[t]
\centering
\includegraphics[width=8.5cm]{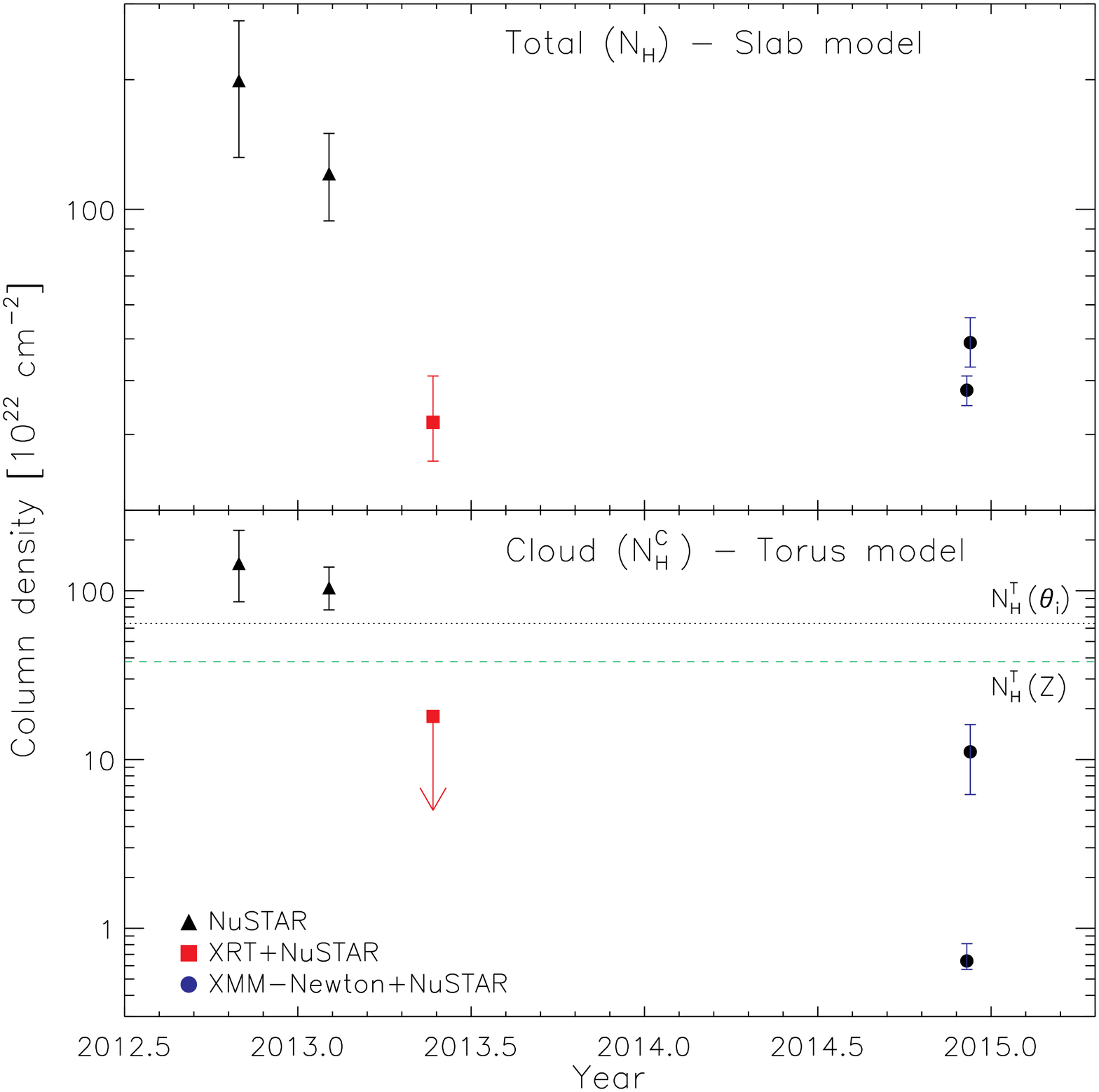}
\label{fig:NHvstime}
\caption{Column density variability of IC\,751. {\it Top panel}: total line-of-sight column density ($N_{\rm H}$) obtained with the slab model described in Sect\,\ref{sect:specAnalysis}. {\it Bottom panel}: column density of the varying absorber (cloud column density, $N_{\rm\,H}^{\rm\,C}$, see Fig.\,\ref{fig:physicalGeom}) obtained with the torus spectral model discussed in Sect.\,\ref{sect:physical modelling}. The black dotted line and the green dashed line represent the values of the non-variable column density of the torus obtained by using \textsc{MYTorus} in its standard [$N_{\rm\,H}^{\rm\,T}(\theta_{\rm\,i})$] and decoupled [$N_{\rm\,H}^{\rm\,T}(Z)$] mode, respectively. The values of the line-of-sight column density are those obtained by using \textsc{MYTorus} in the decoupled mode (Sect.\,\ref{sect:mytdec}).}
\end{figure}

The X-ray observations of IC\,751 presented here show clear evidence of changes in the line-of-sight column density, with the X-ray source being obscured by Compton-thick [$\log (N_{\rm H}/\rm\,cm^{-2})\sim 24.3$] material in two observations, and by Compton-thin material [$\log (N_{\rm H}/\rm\,cm^{-2})\simeq 23.60$] in three observations. This result is confirmed adopting both the slab (Sect.\,\ref{sect:specAnalysis}, top panel of Fig.\,\ref{fig:NHvstime}) and the torus (Sect.\,\ref{sect:physical modelling}, bottom panel of Fig.\,\ref{fig:NHvstime}) X-ray spectral models. In particular, by using a physical torus model we were able, assuming a smooth and azimuthally symmetric torus, to disentangle the intrinsic obscuration associated with a non-varying absorber from the column density of the varying absorber. Variations in the observed line-of-sight column density might be related either to: i) intrinsic variation of the absorbing material, caused by moving clouds;  ii) changes in the intensity of the nuclear radiation, which would cause a variation in the ionization state of the absorbing material. Although the observations during which the highest values of the column density were found also correspond to the stage in which the source was more luminous, for a low-density photo-ionised absorber there might be delay between the flux variation and the response of the absorber, so that the second scenario cannot be completely discarded. In the following we will however assume that changes in $N_{\rm\,H}$ are related to clouds eclipsing the X-ray source, as found for several other changing-look AGN (e.g., \citealp{Risaliti:2005kl}).

Following \cite{Risaliti:2007qa} and \cite{Marinucci:2013pi}, the distance of the cloud from the X-ray source ($R_c$) can be estimated by considering that the size of the source and that of the cloud are similar ($D_{c}\simeq D_{\rm\,s}$), and that the transverse velocity is given by the ratio between the size of the source ($D_{s}$) and the crossing time $T_{cr}$: $V_{k}=D_{s}/T_{cr}$. It must be remarked that in Obs.\,2 we found possible evidence of partial covering, which would imply that $D_{c} < D_{s}$. By applying the slab model we found that the cloud covers 83\% of the X-ray source. The value of the covering factor was larger when adopting the torus model ($\sim 96\%$). Given the rather large values of the covering factor, taking the possible difference between $D_{c}$ and $D_{s}$ into account does not significantly affect our results. Assuming that the cloud is moving with a Keplerian velocity, we obtain:
\begin{equation}
R_c=\frac{GM_{\rm\,BH}}{V_k^2}=\frac{GM_{\rm\,BH}T_{cr}^2}{D_{s}^2}.
\end{equation}
Micro-lensing (e.g., \citealp{Chartas:2002qf,Chartas:2009bh}), occultation studies \citep{Risaliti:2009mi} and large-amplitude rapid X-ray variability have shown that the size of the X-ray source is $D_{\rm\,s}\simeq 10\,r_{\rm\,g}$, where $r_{\rm\,g}=GM_{\rm\,BH}/c^2$. Assuming that $D_{\rm\,s}= 10\,r_{\rm\,g}$, we obtain

\begin{equation}\label{eq:radius}
R_c=\frac{GM_{\rm\,BH}T^2_{\rm\,cr}}{10^2 R_{\rm\,G}^2}\simeq 2\rm\,pc\,M_{8.5}\,R_{10}^{-2}\,T_{10}^2,
\end{equation}

\noindent where $T_{10}$ is the crossing time in units of ten days ($8.64\times 10^5$\,s) and $M_{8.5}=M_{\rm\,BH}/10^{8.5}M_{\odot}$.

The black hole mass of IC\,751 has been recently obtained by the study of the stellar velocity dispersion, as part of work aimed at constraining the characteristics of {\it Swift}/BAT selected AGN in the optical band (Koss et al. in prep.), and is $\log (M_{\rm\,BH}/ M_{\odot})\simeq 8.5$. By using Eq.\,\ref{eq:radius}, and considering that i) no significant variation was found during the 50\,ks of {\it NuSTAR} Obs.\,3  (over a total of 100\,ks), and that ii) the shortest interval in which a variation of the CT material is evident is between Obs.\,3 and Obs.\,4, which were carried out 108\,days apart, we can say that the distance of the cloud is between $R_{\rm\,min}=0.027$\,pc ($\sim 32$\,light-days) and $R_{\rm\,max} \simeq 230$\,pc.

Optical reverberation-mapping studies have shown that the radius of the BLR scales with the square root of the luminosity (e.g., \citealp{Kaspi:2005ys}).
According to \citet{Kaspi:2005ys}, considering the H$\beta$ lags and the results obtained averaging different observations of the same object, the radius of the BLR is given by
\begin{equation}
\frac{R_{\rm BLR}}{10 \rm\,lt-days}=0.86\times \left(\frac{L_{2-10}}{10^{43}\rm\,erg\,s^{-1}}\right)^{0.532}.
\end{equation}
For the average 2--10\,keV luminosity of IC\,751 we find $R_{\rm BLR}=11.7$\,light-days, which implies that the obscuring clouds responsible for the variation of $N_{\rm\,H}$ are beyond the emission-weighted average radius of the BLR. Using a similar approach it is possible to put constraints on the location of the hot inner wall of the torus, which is also known to scale with the square-root of the luminosity (e.g., \citealp{Suganuma:2006zr}, \citealp{Kishimoto:2011uq}). Following \cite{Tristram:2011ve} (Fig.\,4 of their paper) the inner radius of the hot dust ($R_{\rm NIR}$), obtained from $K$-band reverberation, can be approximated by:
\begin{equation}\label{Eq:tor1}
\log \frac{R_{\rm NIR}}{1\rm\,pc}= -23.10+0.5\log L_{14-195},
\end{equation}
where $L_{14-195}$ is the 14--195\,keV luminosity (in $\rm\,erg\,s^{-1}$). At 12$\mu$m, interferometric studies (\citealp{Tristram:2011ve}, see also \citealp{Burtscher:2013zt}) have shown that the size of the mid-IR emitting region ($R_{\rm MIR}$) in AGN can be estimated by
\begin{equation}\label{Eq:tor2}
\log \frac{R_{\rm MIR}}{1\rm\,pc}= -21.62+0.5\log L_{14-195}.
\vspace{0.2mm}
\end{equation}
The 70-month averaged 14--195\,keV luminosity of IC\,751 is $\log (L_{14-195}/\rm\,erg\,s^{-1})=43.47$ \citep{Baumgartner:2013uq}, which corresponds to $R_{\rm NIR}=0.04$\,pc ($\sim 48$\,light-days) and $R_{\rm MIR}=1.3$\,pc. From this we can conclude that the absorbing material could be related to the outer BLR, to clumps in the molecular torus or even to material located at further distance from the SMBH.

A change in the Compton-thin absorber is also found between Obs.\,5 and Obs.\,6 using the slab spectral model and \textsc{MyTORUS} in its decoupled mode. The two observations were carried out about 48\,hours apart, which means (applying Eq.\,\ref{eq:radius}) that for the Compton-thin material the $R\lesssim R_{\rm\,max}=0.08$\,pc ($\sim 95$\,light-days). If the Compton-thin and CT clouds are located at the same distance from the X-ray source, then the regions where the varying absorbers are located is between 32 and 95\,light-days. This would imply that the absorber is consistent with being located either in the BLR or in the inner side of the dusty torus. Recent work has shown that the Fe K$\alpha$ might also arise in this region \citep{Minezaki:2015vn,Gandhi:2015cr} Assuming that the cloud has about the same size of the X-ray source, i.e. $\sim 10\,r_{\rm\,g}$ ($\sim 4.7\times 10^{14}$\,cm), and that its column density is $1.5\times 10^{24}\rm\,cm^{-2}$, the density of the cloud would be $n\sim 3.2\times 10^9 \rm cm^{-3}$. This value is in agreement with that expected for the BLR clouds (e.g., \citealp{Peterson:1997nx}). A BLR origin for the varying absorber in IC\,751 would fit what has been found so far for other changing-look AGN, several of which show absorbers compatible with being part of the BLR (e.g., \citealp{Maiolino:2010fu,Risaliti:2010ve,Burtscher:2016fk}). 

Thanks to its broad-band coverage, {\it NuSTAR} is a very powerful tool to study obscuration in AGN, and it has been shown to be fundamental to well constrain the line-of-sight column density (e.g., \citealp{Arevalo:2014nx,Gandhi:2014bh,Koss:2015qf,Annuar:2015ve,Bauer:2014cr,Lansbury:2015vn}). Studying type-II quasars, \citet{Lansbury:2015vn} have shown that the estimates of $N_{\rm\,H}$ obtained by {\it NuSTAR} are 2.5--1600 times higher than previous constraints from {\it XMM-Newton} and {\it Chandra}. This shows that, in the absence of high-quality broad-band observations, it would be possible to miss changing-look events for weak sources. Another clear example is given by the recent detection of an unveiling event in NGC\,1068 \citep{Marinucci:2015kx}, which would have been missed by observations carried out below 10\,keV. Repeated {\it NuSTAR} observations of obscured sources might therefore uncover a significant number of new changing-look events. \cite{Burtscher:2016fk} have recently reanalysed the relation between $N_{\rm\,H}$ and the optical obscuration $A_{\rm\,V}$, and found that in several cases the deviation of $N_{\rm\,H}/A_{\rm\,V}$ from the Galactic value is due to variable absorption. This would imply that ideal targets to study occultations of the X-ray source are objects showing a large deviation from the Galactic $N_{\rm\,H}/A_{\rm\,V}$ value.

\medskip
\medskip
\medskip
\medskip

\section{Summary and conclusions}

We reported here on the spectral analysis of five {\it NuSTAR} observations of the type-2 AGN IC\,751, three of which were combined with {\it XMM-Newton} or {\it Swift}/XRT observations in the 0.3--10\,keV range. IC\,751 is the first changing-look AGN (i.e. an object that has been observed both in a Compton-thin and a CT state) discovered by {\it NuSTAR}. We find that the X-ray source was obscured by CT material during the first two observations, while its line of sight obscuration is found to be Compton-thin during the following observations, which implies that absorption varies on timescales of $\lesssim 3$\,months. Changes of the line-of-sight column density are also found on a time-scale of $\sim 48$\,hours ($\Delta N_{\rm\,H}\sim 10^{23}\rm\,cm^{-2}$). While we cannot constrain the location of the absorber precisely, by considering the lack of spectral variability during the longest {\it NuSTAR} observation we can infer the minimum distance to be further than the emission-weighted average radius of the BLR. Assuming that the varying Compton-thin and CT clouds are located at the same distance from the X-ray source, then the material is located between 32 and 95\,light-days. The absorber could therefore be related either to the external part of the BLR or to the inner part of the dusty torus, although the BLR origin might be slightly favored since the density of the clouds is found to be consistent with the value expected for BLR clouds. By adopting a physical torus X-ray spectral model, we are able to disentangle the column density of the non-varying absorber ($N_{\rm\,H}\sim 3.8\times 10^{23}\rm\,cm^{-2}$) from that of the varying clouds [$N_{\rm\,H}\sim(1-150)\times10^{22}\rm\,cm^{-2}$], and to put constraints on the column density of the reprocessing material ($N_{\rm\,H} \sim 6 \times 10^{24}\rm\,cm^{-2}$). We found that the X-ray source is highly variable both in the 2--10\,keV and 10--50\,keV bands. Future observational campaigns on IC\,751 in the X-ray band will be able to improve the constraints on the location of the varying absorber, and confirm or not whether it is related to clouds in the BLR as it has been found for several objects of this class.

\acknowledgments
We thank the anonymous referee for his/her comments, that helped us to improve the quality of our manuscript. CR acknowledges Marko Stalevski and Sebastien Guillot for fruitful discussion, and Chin-Shin Chang for her comments on the manuscript. This research has made use of the NuSTAR Data Analysis Software (\textsc{NuSTARDAS}) jointly developed by the ASI Science Data Center (ASDC, Italy) and the California Institute of Technology (Caltech, USA).
We acknowledge financial support from the CONICYT-Chile grants ''EMBIGGEN" Anillo ACT1101 (CR, ET, FEB, PA), FONDECYT 1141218 (CR, FEB), FONDECYT 1140304 (PA), Basal-CATA PFB--06/2007 (CR, ET, FEB), {\it NuSTAR} subcontract 44A--1092750 (WNB), the Swiss National Science Foundation and Ambizione fellowship grant PZ00P2\textunderscore154799/1 (MK) and the Ministry of Economy, Development, and Tourism's Millennium Science Initiative through grant IC120009, awarded to The Millennium Institute of Astrophysics, MAS (FEB). 
This research has made use of the NASA/IPAC Extragalactic Database (NED) which is operated by the Jet Propulsion Laboratory, of data obtained from the High Energy Astrophysics Science Archive Research Center (HEASARC), provided by NASA's Goddard Space Flight Center, and of the SIMBAD Astronomical Database which is operated by the Centre de Donn\'ees astronomiques de Strasbourg.

{\it Facilities:} \facility{NuSTAR}, \facility{Swift}, \facility{XMM-Newton}.

\smallskip

\bibliographystyle{apj} 
\bibliography{IC0751_ref.bib}

\begin{thebibliography}{82}
\expandafter\ifx\csname natexlab\endcsname\relax\def\natexlab#1{#1}\fi

\bibitem[{{Annuar} {et~al.}(2015){Annuar}, {Gandhi}, {Alexander}, {Lansbury},
  {Ar{\'e}valo}, {Ballantyne}, {Balokovi{\'c}}, {Bauer}, {Boggs}, {Brandt},
  {Brightman}, {Christensen}, {Craig}, {Del Moro}, {Hailey}, {Harrison},
  {Hickox}, {Matt}, {Puccetti}, {Ricci}, {Rigby}, {Stern}, {Walton},
  {Zappacosta}, \& {Zhang}}]{Annuar:2015ve}
{Annuar}, A., {Gandhi}, P., {Alexander}, D.~M., {et~al.} 2015, ArXiv e-prints

\bibitem[{{Ar{\'e}valo} {et~al.}(2014){Ar{\'e}valo}, {Bauer}, {Puccetti},
  {Walton}, {Koss}, {Boggs}, {Brandt}, {Brightman}, {Christensen}, {Comastri},
  {Craig}, {Fuerst}, {Gandhi}, {Grefenstette}, {Hailey}, {Harrison}, {Luo},
  {Madejski}, {Madsen}, {Marinucci}, {Matt}, {Saez}, {Stern}, {Stuhlinger},
  {Treister}, {Urry}, \& {Zhang}}]{Arevalo:2014nx}
{Ar{\'e}valo}, P., {Bauer}, F.~E., {Puccetti}, S., {et~al.} 2014, \apj, 791, 81

\bibitem[{{Arnaud}(1996)}]{Arnaud:1996kx}
{Arnaud}, K.~A. 1996, in Astronomical Society of the Pacific Conference Series,
  Vol. 101, Astronomical Data Analysis Software and Systems V, ed.
  {G.~H.~Jacoby \& J.~Barnes}, 17

\bibitem[{{Asmus} {et~al.}(2015){Asmus}, {Gandhi}, {Hoenig}, {Smette}, \&
  {Duschl}}]{Asmus:2015ly}
{Asmus}, D., {Gandhi}, P., {Hoenig}, S.~F., {Smette}, A., \& {Duschl}, W.~J.
  2015, ArXiv e-prints

\bibitem[{{Bauer} {et~al.}(2014){Bauer}, {Arevalo}, {Walton}, {Koss},
  {Puccetti}, {Gandhi}, {Stern}, {Alexander}, {Balokovic}, {Boggs}, {Brandt},
  {Brightman}, {Christensen}, {Comastri}, {Craig}, {Del Moro}, {Hailey},
  {Harrison}, {Hickox}, {Luo}, {Markwardt}, {Marinucci}, {Matt}, {Rigby},
  {Rivers}, {Saez}, {Treister}, {Urry}, \& {Zhang}}]{Bauer:2014cr}
{Bauer}, F.~E., {Arevalo}, P., {Walton}, D.~J., {et~al.} 2014, ArXiv e-prints

\bibitem[{{Baumgartner} {et~al.}(2013){Baumgartner}, {Tueller}, {Markwardt},
  {Skinner}, {Barthelmy}, {Mushotzky}, {Evans}, \&
  {Gehrels}}]{Baumgartner:2013uq}
{Baumgartner}, W.~H., {Tueller}, J., {Markwardt}, C.~B., {et~al.} 2013, \apjs,
  207, 19

\bibitem[{{Beckmann} {et~al.}(2011){Beckmann}, {Jean}, {Lubi{\'n}ski}, {Soldi},
  \& {Terrier}}]{Beckmann:2011hc}
{Beckmann}, V., {Jean}, P., {Lubi{\'n}ski}, P., {Soldi}, S., \& {Terrier}, R.
  2011, \aap, 531, A70

\bibitem[{{Bianchi} {et~al.}(2005){Bianchi}, {Guainazzi}, {Matt}, {Chiaberge},
  {Iwasawa}, {Fiore}, \& {Maiolino}}]{Bianchi:2005ys}
{Bianchi}, S., {Guainazzi}, M., {Matt}, G., {et~al.} 2005, \aap, 442, 185

\bibitem[{{Bianchi} {et~al.}(2012){Bianchi}, {Maiolino}, \&
  {Risaliti}}]{Bianchi:2012cr}
{Bianchi}, S., {Maiolino}, R., \& {Risaliti}, G. 2012, Advances in Astronomy,
  2012, 17

\bibitem[{{Bianchi} {et~al.}(2009){Bianchi}, {Piconcelli}, {Chiaberge},
  {Bail{\'o}n}, {Matt}, \& {Fiore}}]{Bianchi:2009ly}
{Bianchi}, S., {Piconcelli}, E., {Chiaberge}, M., {et~al.} 2009, \apj, 695, 781

\bibitem[{{Braito} {et~al.}(2013){Braito}, {Ballo}, {Reeves}, {Risaliti},
  {Ptak}, \& {Turner}}]{Braito:2013nx}
{Braito}, V., {Ballo}, L., {Reeves}, J.~N., {et~al.} 2013, \mnras, 428, 2516

\bibitem[{{Burrows} {et~al.}(2005){Burrows}, {Hill}, {Nousek}, {Kennea},
  {Wells}, {Osborne}, {Abbey}, {Beardmore}, {Mukerjee}, {Short}, {Chincarini},
  {Campana}, {Citterio}, {Moretti}, {Pagani}, {Tagliaferri}, {Giommi},
  {Capalbi}, {Tamburelli}, {Angelini}, {Cusumano}, {Br{\"a}uninger}, {Burkert},
  \& {Hartner}}]{Burrows:2005vn}
{Burrows}, D.~N., {Hill}, J.~E., {Nousek}, J.~A., {et~al.} 2005, \ssr, 120, 165

\bibitem[{{Burtscher} {et~al.}(2013){Burtscher}, {Meisenheimer}, {Tristram},
  {Jaffe}, {H{\"o}nig}, {Davies}, {Kishimoto}, {Pott}, {R{\"o}ttgering},
  {Schartmann}, {Weigelt}, \& {Wolf}}]{Burtscher:2013zt}
{Burtscher}, L., {Meisenheimer}, K., {Tristram}, K.~R.~W., {et~al.} 2013, \aap,
  558, A149

\bibitem[{{Burtscher} {et~al.}(2016){Burtscher}, {Davies}, {Graci{\'a}-Carpio},
  {Koss}, {Lin}, {Lutz}, {Nandra}, {Netzer}, {Orban de Xivry}, {Ricci},
  {Rosario}, {Veilleux}, {Contursi}, {Genzel}, {Schnorr-M{\"u}ller},
  {Sternberg}, {Sturm}, \& {Tacconi}}]{Burtscher:2016fk}
{Burtscher}, L., {Davies}, R.~I., {Graci{\'a}-Carpio}, J., {et~al.} 2016, \aap,
  586, A28

\bibitem[{{Cash}(1979)}]{Cash:1979fk}
{Cash}, W. 1979, \apj, 228, 939

\bibitem[{{Chartas} {et~al.}(2002){Chartas}, {Agol}, {Eracleous}, {Garmire},
  {Bautz}, \& {Morgan}}]{Chartas:2002qf}
{Chartas}, G., {Agol}, E., {Eracleous}, M., {et~al.} 2002, \apj, 568, 509

\bibitem[{{Chartas} {et~al.}(2009){Chartas}, {Kochanek}, {Dai}, {Poindexter},
  \& {Garmire}}]{Chartas:2009bh}
{Chartas}, G., {Kochanek}, C.~S., {Dai}, X., {Poindexter}, S., \& {Garmire}, G.
  2009, \apj, 693, 174

\bibitem[{{de Vaucouleurs} {et~al.}(1991){de Vaucouleurs}, {de Vaucouleurs},
  {Corwin}, {Buta}, {Paturel}, \& {Fouqu{\'e}}}]{de-Vaucouleurs:1991nx}
{de Vaucouleurs}, G., {de Vaucouleurs}, A., {Corwin}, Jr., H.~G., {et~al.}
  1991, {Third Reference Catalogue of Bright Galaxies. Volume I: Explanations
  and references. Volume II: Data for galaxies between 0$^{h}$ and 12$^{h}$.
  Volume III: Data for galaxies between 12$^{h}$ and 24$^{h}$.}

\bibitem[{{Elvis} {et~al.}(2004){Elvis}, {Risaliti}, {Nicastro}, {Miller},
  {Fiore}, \& {Puccetti}}]{Elvis:2004vn}
{Elvis}, M., {Risaliti}, G., {Nicastro}, F., {et~al.} 2004, \apjl, 615, L25

\bibitem[{{Falco} {et~al.}(1999){Falco}, {Kurtz}, {Geller}, {Huchra}, {Peters},
  {Berlind}, {Mink}, {Tokarz}, \& {Elwell}}]{Falco:1999fk}
{Falco}, E.~E., {Kurtz}, M.~J., {Geller}, M.~J., {et~al.} 1999, \pasp, 111, 438

\bibitem[{{Gabriel} {et~al.}(2004){Gabriel}, {Denby}, {Fyfe}, {Hoar}, {Ibarra},
  {Ojero}, {Osborne}, {Saxton}, {Lammers}, \& {Vacanti}}]{Gabriel:2004fk}
{Gabriel}, C., {Denby}, M., {Fyfe}, D.~J., {et~al.} 2004, in Astronomical
  Society of the Pacific Conference Series, Vol. 314, Astronomical Data
  Analysis Software and Systems (ADASS) XIII, ed. F.~{Ochsenbein}, M.~G.
  {Allen}, \& D.~{Egret}, 759

\bibitem[{{Gallagher} {et~al.}(2004){Gallagher}, {Brandt}, {Wills}, {Charlton},
  {Chartas}, \& {Laor}}]{Gallagher:2004uq}
{Gallagher}, S.~C., {Brandt}, W.~N., {Wills}, B.~J., {et~al.} 2004, \apj, 603,
  425

\bibitem[{{Gandhi} {et~al.}(2015){Gandhi}, {Hoenig}, \&
  {Kishimoto}}]{Gandhi:2015cr}
{Gandhi}, P., {Hoenig}, S.~F., \& {Kishimoto}, M. 2015, ArXiv e-prints

\bibitem[{{Gandhi} {et~al.}(2009){Gandhi}, {Horst}, {Smette}, {H{\"o}nig},
  {Comastri}, {Gilli}, {Vignali}, \& {Duschl}}]{Gandhi:2009zr}
{Gandhi}, P., {Horst}, H., {Smette}, A., {et~al.} 2009, \aap, 502, 457

\bibitem[{{Gandhi} {et~al.}(2014){Gandhi}, {Lansbury}, {Alexander}, {Stern},
  {Ar{\'e}valo}, {Ballantyne}, {Balokovi{\'c}}, {Bauer}, {Boggs}, {Brandt},
  {Brightman}, {Christensen}, {Comastri}, {Craig}, {Del Moro}, {Elvis},
  {Fabian}, {Hailey}, {Harrison}, {Hickox}, {Koss}, {LaMassa}, {Luo},
  {Madejski}, {Ptak}, {Puccetti}, {Teng}, {Urry}, {Walton}, \&
  {Zhang}}]{Gandhi:2014bh}
{Gandhi}, P., {Lansbury}, G.~B., {Alexander}, D.~M., {et~al.} 2014, \apj, 792,
  117

\bibitem[{{Gehrels} {et~al.}(2004){Gehrels}, {Chincarini}, {Giommi}, {Mason},
  {Nousek}, {Wells}, {White}, {Barthelmy}, {Burrows}, {Cominsky}, {Hurley},
  {Marshall}, {M{\'e}sz{\'a}ros}, {Roming}, {Angelini}, {Barbier}, {Belloni},
  {Campana}, {Caraveo}, {Chester}, {Citterio}, {Cline}, {Cropper}, {Cummings},
  {Dean}, {Feigelson}, {Fenimore}, {Frail}, {Fruchter}, {Garmire}, {Gendreau},
  {Ghisellini}, {Greiner}, {Hill}, {Hunsberger}, {Krimm}, {Kulkarni}, {Kumar},
  {Lebrun}, {Lloyd-Ronning}, {Markwardt}, {Mattson}, {Mushotzky}, {Norris},
  {Osborne}, {Paczynski}, {Palmer}, {Park}, {Parsons}, {Paul}, {Rees},
  {Reynolds}, {Rhoads}, {Sasseen}, {Schaefer}, {Short}, {Smale}, {Smith},
  {Stella}, {Tagliaferri}, {Takahashi}, {Tashiro}, {Townsley}, {Tueller},
  {Turner}, {Vietri}, {Voges}, {Ward}, {Willingale}, {Zerbi}, \&
  {Zhang}}]{Gehrels:2004dq}
{Gehrels}, N., {Chincarini}, G., {Giommi}, P., {et~al.} 2004, \apj, 611, 1005

\bibitem[{{Guainazzi}(2002)}]{Guainazzi:2002dz}
{Guainazzi}, M. 2002, \mnras, 329, L13

\bibitem[{{Guainazzi} {et~al.}(2002){Guainazzi}, {Matt}, {Fiore}, \&
  {Perola}}]{Guainazzi:2002fv}
{Guainazzi}, M., {Matt}, G., {Fiore}, F., \& {Perola}, G.~C. 2002, \aap, 388,
  787

\bibitem[{{Harrison} {et~al.}(2013){Harrison}, {Craig}, {Christensen},
  {Hailey}, {Zhang}, {Boggs}, {Stern}, {Cook}, {Forster}, {Giommi},
  {Grefenstette}, {Kim}, {Kitaguchi}, {Koglin}, {Madsen}, {Mao}, {Miyasaka},
  {Mori}, {Perri}, {Pivovaroff}, {Puccetti}, {Rana}, {Westergaard}, {Willis},
  {Zoglauer}, {An}, {Bachetti}, {Barri{\`e}re}, {Bellm}, {Bhalerao},
  {Brejnholt}, {Fuerst}, {Liebe}, {Markwardt}, {Nynka}, {Vogel}, {Walton},
  {Wik}, {Alexander}, {Cominsky}, {Hornschemeier}, {Hornstrup}, {Kaspi},
  {Madejski}, {Matt}, {Molendi}, {Smith}, {Tomsick}, {Ajello}, {Ballantyne},
  {Balokovi{\'c}}, {Barret}, {Bauer}, {Blandford}, {Brandt}, {Brenneman},
  {Chiang}, {Chakrabarty}, {Chenevez}, {Comastri}, {Dufour}, {Elvis}, {Fabian},
  {Farrah}, {Fryer}, {Gotthelf}, {Grindlay}, {Helfand}, {Krivonos}, {Meier},
  {Miller}, {Natalucci}, {Ogle}, {Ofek}, {Ptak}, {Reynolds}, {Rigby},
  {Tagliaferri}, {Thorsett}, {Treister}, \& {Urry}}]{Harrison:2013uq}
{Harrison}, F.~A., {Craig}, W.~W., {Christensen}, F.~E., {et~al.} 2013, \apj,
  770, 103

\bibitem[{{Immler} {et~al.}(2003){Immler}, {Brandt}, {Vignali}, {Bauer},
  {Crenshaw}, {Feldmeier}, \& {Kraemer}}]{Immler:2003fk}
{Immler}, S., {Brandt}, W.~N., {Vignali}, C., {et~al.} 2003, \aj, 126, 153

\bibitem[{{Jansen} {et~al.}(2001){Jansen}, {Lumb}, {Altieri}, {Clavel}, {Ehle},
  {Erd}, {Gabriel}, {Guainazzi}, {Gondoin}, {Much}, {Munoz}, {Santos},
  {Schartel}, {Texier}, \& {Vacanti}}]{Jansen:2001vn}
{Jansen}, F., {Lumb}, D., {Altieri}, B., {et~al.} 2001, \aap, 365, L1

\bibitem[{{Kalberla} {et~al.}(2005){Kalberla}, {Burton}, {Hartmann}, {Arnal},
  {Bajaja}, {Morras}, \& {P{\"o}ppel}}]{Kalberla:2005fk}
{Kalberla}, P.~M.~W., {Burton}, W.~B., {Hartmann}, D., {et~al.} 2005, \aap,
  440, 775

\bibitem[{{Kaspi} {et~al.}(2005){Kaspi}, {Maoz}, {Netzer}, {Peterson},
  {Vestergaard}, \& {Jannuzi}}]{Kaspi:2005ys}
{Kaspi}, S., {Maoz}, D., {Netzer}, H., {et~al.} 2005, \apj, 629, 61

\bibitem[{{Kishimoto} {et~al.}(2011){Kishimoto}, {H{\"o}nig}, {Antonucci},
  {Barvainis}, {Kotani}, {Tristram}, {Weigelt}, \& {Levin}}]{Kishimoto:2011uq}
{Kishimoto}, M., {H{\"o}nig}, S.~F., {Antonucci}, R., {et~al.} 2011, \aap, 527,
  A121

\bibitem[{{Koss} {et~al.}(2015){Koss}, {Romero-Canizales}, {Baronchelli},
  {Teng}, {Balokovic}, {Puccetti}, {Bauer}, {Arevalo}, {Assef}, {Ballantyne},
  {Brandt}, {Brightman}, {Comastri}, {Gandhi}, {Harrison}, {Luo}, {Schawinski},
  {Stern}, \& {Treister}}]{Koss:2015qf}
{Koss}, M.~J., {Romero-Canizales}, C., {Baronchelli}, L., {et~al.} 2015, ArXiv
  e-prints

\bibitem[{{LaMassa} {et~al.}(2015){LaMassa}, {Cales}, {Moran}, {Myers},
  {Richards}, {Eracleous}, {Heckman}, {Gallo}, \& {Urry}}]{LaMassa:2015ys}
{LaMassa}, S.~M., {Cales}, S., {Moran}, E.~C., {et~al.} 2015, \apj, 800, 144

\bibitem[{{Lamer} {et~al.}(2003){Lamer}, {Uttley}, \& {McHardy}}]{Lamer:2003oq}
{Lamer}, G., {Uttley}, P., \& {McHardy}, I.~M. 2003, \mnras, 342, L41

\bibitem[{{Lansbury} {et~al.}(2015){Lansbury}, {Gandhi}, {Alexander}, {Assef},
  {Aird}, {Annuar}, {Ballantyne}, {Balokovic}, {Bauer}, {Boggs}, {Brandt},
  {Brightman}, {Christensen}, {Civano}, {Comastri}, {Craig}, {Del Moro},
  {Grefenstette}, {Hailey}, {Harrison}, {Hickox}, {Koss}, {LaMassa}, {Luo},
  {Puccetti}, {Stern}, {Treister}, {Vignali}, {Zappacosta}, \&
  {Zhang}}]{Lansbury:2015vn}
{Lansbury}, G.~B., {Gandhi}, P., {Alexander}, D.~M., {et~al.} 2015, ArXiv
  e-prints

\bibitem[{{Longinotti} {et~al.}(2009){Longinotti}, {Bianchi}, {Ballo}, {de La
  Calle}, \& {Guainazzi}}]{Longinotti:2009tg}
{Longinotti}, A.~L., {Bianchi}, S., {Ballo}, L., {de La Calle}, I., \&
  {Guainazzi}, M. 2009, \mnras, 394, L1

\bibitem[{{Madsen} {et~al.}(2015){Madsen}, {Harrison}, {Markwardt}, {An},
  {Grefenstette}, {Bachetti}, {Miyasaka}, {Kitaguchi}, {Bhalerao},
  {Christensen}, {Craig}, {Fuerst}, {Walton}, {Hailey}, {Rana}, {Stern},
  {Westergaard}, \& {Zhang}}]{Madsen:2015uq}
{Madsen}, K.~K., {Harrison}, F.~A., {Markwardt}, C., {et~al.} 2015, ArXiv
  e-prints

\bibitem[{{Magdziarz} \& {Zdziarski}(1995)}]{Magdziarz:1995pi}
{Magdziarz}, P., \& {Zdziarski}, A.~A. 1995, \mnras, 273, 837

\bibitem[{{Maiolino} {et~al.}(2010){Maiolino}, {Risaliti}, {Salvati},
  {Pietrini}, {Torricelli-Ciamponi}, {Elvis}, {Fabbiano}, {Braito}, \&
  {Reeves}}]{Maiolino:2010fu}
{Maiolino}, R., {Risaliti}, G., {Salvati}, M., {et~al.} 2010, \aap, 517, A47

\bibitem[{{Marchese} {et~al.}(2012){Marchese}, {Braito}, {Della Ceca},
  {Caccianiga}, \& {Severgnini}}]{Marchese:2012kx}
{Marchese}, E., {Braito}, V., {Della Ceca}, R., {Caccianiga}, A., \&
  {Severgnini}, P. 2012, \mnras, 421, 1803

\bibitem[{{Marinucci} {et~al.}(2013){Marinucci}, {Risaliti}, {Wang}, {Bianchi},
  {Elvis}, {Matt}, {Nardini}, \& {Braito}}]{Marinucci:2013pi}
{Marinucci}, A., {Risaliti}, G., {Wang}, J., {et~al.} 2013, \mnras, 429, 2581

\bibitem[{{Marinucci} {et~al.}(2015){Marinucci}, {Bianchi}, {Matt},
  {Alexander}, {Balokovic}, {Bauer}, {Brandt}, {Gandhi}, {Guainazzi},
  {Harrison}, {Iwasawa}, {Koss}, {Madsen}, {Nicastro}, {Puccetti}, {Ricci},
  {Stern}, \& {Walton}}]{Marinucci:2015kx}
{Marinucci}, A., {Bianchi}, S., {Matt}, G., {et~al.} 2015, ArXiv e-prints

\bibitem[{{Markowitz} {et~al.}(2014){Markowitz}, {Krumpe}, \&
  {Nikutta}}]{Markowitz:2014oq}
{Markowitz}, A.~G., {Krumpe}, M., \& {Nikutta}, R. 2014, \mnras, 439, 1403

\bibitem[{{Matt} {et~al.}(2003){Matt}, {Guainazzi}, \&
  {Maiolino}}]{Matt:2003vn}
{Matt}, G., {Guainazzi}, M., \& {Maiolino}, R. 2003, \mnras, 342, 422

\bibitem[{{Minezaki} \& {Matsushita}(2015)}]{Minezaki:2015vn}
{Minezaki}, T., \& {Matsushita}, K. 2015, \apj, 802, 98

\bibitem[{{Miniutti} {et~al.}(2014){Miniutti}, {Sanfrutos}, {Beuchert},
  {Ag{\'{\i}}s-Gonz{\'a}lez}, {Longinotti}, {Piconcelli}, {Krongold},
  {Guainazzi}, {Bianchi}, {Matt}, \&
  {Jim{\'e}nez-Bail{\'o}n}}]{Miniutti:2014nx}
{Miniutti}, G., {Sanfrutos}, M., {Beuchert}, T., {et~al.} 2014, \mnras, 437,
  1776

\bibitem[{{Murphy} \& {Yaqoob}(2009)}]{Murphy:2009ly}
{Murphy}, K.~D., \& {Yaqoob}, T. 2009, \mnras, 397, 1549

\bibitem[{{Nardini} \& {Risaliti}(2011)}]{Nardini:2011ij}
{Nardini}, E., \& {Risaliti}, G. 2011, \mnras, 417, 2571

\bibitem[{{Peterson}(1997)}]{Peterson:1997nx}
{Peterson}, B.~M. 1997, {An Introduction to Active Galactic Nuclei}

\bibitem[{{Piconcelli} {et~al.}(2007){Piconcelli}, {Bianchi}, {Guainazzi},
  {Fiore}, \& {Chiaberge}}]{Piconcelli:2007bs}
{Piconcelli}, E., {Bianchi}, S., {Guainazzi}, M., {Fiore}, F., \& {Chiaberge},
  M. 2007, \aap, 466, 855

\bibitem[{{Pounds} {et~al.}(2004){Pounds}, {Reeves}, {Page}, \&
  {O'Brien}}]{Pounds:2004bh}
{Pounds}, K.~A., {Reeves}, J.~N., {Page}, K.~L., \& {O'Brien}, P.~T. 2004,
  \apj, 616, 696

\bibitem[{{Puccetti} {et~al.}(2007){Puccetti}, {Fiore}, {Risaliti}, {Capalbi},
  {Elvis}, \& {Nicastro}}]{Puccetti:2007zr}
{Puccetti}, S., {Fiore}, F., {Risaliti}, G., {et~al.} 2007, \mnras, 377, 607

\bibitem[{{Ricci} {et~al.}(2014{\natexlab{a}}){Ricci}, {Ueda}, {Ichikawa},
  {Paltani}, {Boissay}, {Gandhi}, {Stalevski}, \& {Awaki}}]{Ricci:2014fk}
{Ricci}, C., {Ueda}, Y., {Ichikawa}, K., {et~al.} 2014{\natexlab{a}}, \aap,
  567, A142

\bibitem[{{Ricci} {et~al.}(2015){Ricci}, {Ueda}, {Koss}, {Trakhtenbrot},
  {Bauer}, \& {Gandhi}}]{Ricci:2015fk}
{Ricci}, C., {Ueda}, Y., {Koss}, M.~J., {et~al.} 2015, \apjl, 815, L13

\bibitem[{{Ricci} {et~al.}(2014{\natexlab{b}}){Ricci}, {Ueda}, {Paltani},
  {Ichikawa}, {Gandhi}, \& {Awaki}}]{Ricci:2014fv}
{Ricci}, C., {Ueda}, Y., {Paltani}, S., {et~al.} 2014{\natexlab{b}}, \mnras,
  441, 3622

\bibitem[{{Risaliti} {et~al.}(2010){Risaliti}, {Elvis}, {Bianchi}, \&
  {Matt}}]{Risaliti:2010ve}
{Risaliti}, G., {Elvis}, M., {Bianchi}, S., \& {Matt}, G. 2010, \mnras, 406,
  L20

\bibitem[{{Risaliti} {et~al.}(2005){Risaliti}, {Elvis}, {Fabbiano}, {Baldi}, \&
  {Zezas}}]{Risaliti:2005kl}
{Risaliti}, G., {Elvis}, M., {Fabbiano}, G., {Baldi}, A., \& {Zezas}, A. 2005,
  \apjl, 623, L93

\bibitem[{{Risaliti} {et~al.}(2007){Risaliti}, {Elvis}, {Fabbiano}, {Baldi},
  {Zezas}, \& {Salvati}}]{Risaliti:2007qa}
{Risaliti}, G., {Elvis}, M., {Fabbiano}, G., {et~al.} 2007, \apjl, 659, L111

\bibitem[{{Risaliti} {et~al.}(2002){Risaliti}, {Elvis}, \&
  {Nicastro}}]{Risaliti:2002uq}
{Risaliti}, G., {Elvis}, M., \& {Nicastro}, F. 2002, \apj, 571, 234

\bibitem[{{Risaliti} {et~al.}(2011){Risaliti}, {Nardini}, {Salvati}, {Elvis},
  {Fabbiano}, {Maiolino}, {Pietrini}, \&
  {Torricelli-Ciamponi}}]{Risaliti:2011fk}
{Risaliti}, G., {Nardini}, E., {Salvati}, M., {et~al.} 2011, \mnras, 410, 1027

\bibitem[{{Risaliti} {et~al.}(2009){Risaliti}, {Salvati}, {Elvis}, {Fabbiano},
  {Baldi}, {Bianchi}, {Braito}, {Guainazzi}, {Matt}, {Miniutti}, {Reeves},
  {Soria}, \& {Zezas}}]{Risaliti:2009mi}
{Risaliti}, G., {Salvati}, M., {Elvis}, M., {et~al.} 2009, \mnras, 393, L1

\bibitem[{{Rivers} {et~al.}(2011){Rivers}, {Markowitz}, \&
  {Rothschild}}]{Rivers:2011kl}
{Rivers}, E., {Markowitz}, A., \& {Rothschild}, R. 2011, \apjl, 742, L29

\bibitem[{{Rivers} {et~al.}(2015{\natexlab{a}}){Rivers}, {Risaliti}, {Walton},
  {Harrison}, {Ar{\'e}valo}, {Baur}, {Boggs}, {Brenneman}, {Brightman},
  {Christensen}, {Craig}, {F{\"u}rst}, {Hailey}, {Hickox}, {Marinucci},
  {Reeves}, {Stern}, \& {Zhang}}]{Rivers:2015uq}
{Rivers}, E., {Risaliti}, G., {Walton}, D.~J., {et~al.} 2015{\natexlab{a}},
  \apj, 804, 107

\bibitem[{{Rivers} {et~al.}(2015{\natexlab{b}}){Rivers}, {Balokovi{\'c}},
  {Ar{\'e}valo}, {Bauer}, {Boggs}, {Brandt}, {Brightman}, {Christensen},
  {Craig}, {Gandhi}, {Hailey}, {Harrison}, {Koss}, {Ricci}, {Stern}, {Walton},
  \& {Zhang}}]{Rivers:2015fk}
{Rivers}, E., {Balokovi{\'c}}, M., {Ar{\'e}valo}, P., {et~al.}
  2015{\natexlab{b}}, ArXiv e-prints

\bibitem[{{Sanfrutos} {et~al.}(2013){Sanfrutos}, {Miniutti},
  {Ag{\'{\i}}s-Gonz{\'a}lez}, {Fabian}, {Miller}, {Panessa}, \&
  {Zoghbi}}]{Sanfrutos:2013cr}
{Sanfrutos}, M., {Miniutti}, G., {Ag{\'{\i}}s-Gonz{\'a}lez}, B., {et~al.} 2013,
  \mnras, 436, 1588

\bibitem[{{Shappee} {et~al.}(2014){Shappee}, {Prieto}, {Grupe}, {Kochanek},
  {Stanek}, {De Rosa}, {Mathur}, {Zu}, {Peterson}, {Pogge}, {Komossa}, {Im},
  {Jencson}, {Holoien}, {Basu}, {Beacom}, {Szczygie{\l}}, {Brimacombe},
  {Adams}, {Campillay}, {Choi}, {Contreras}, {Dietrich}, {Dubberley},
  {Elphick}, {Foale}, {Giustini}, {Gonzalez}, {Hawkins}, {Howell}, {Hsiao},
  {Koss}, {Leighly}, {Morrell}, {Mudd}, {Mullins}, {Nugent}, {Parrent},
  {Phillips}, {Pojmanski}, {Rosing}, {Ross}, {Sand}, {Terndrup}, {Valenti},
  {Walker}, \& {Yoon}}]{Shappee:2014kx}
{Shappee}, B.~J., {Prieto}, J.~L., {Grupe}, D., {et~al.} 2014, \apj, 788, 48

\bibitem[{{Shu} {et~al.}(2010){Shu}, {Yaqoob}, \& {Wang}}]{Shu:2010zr}
{Shu}, X.~W., {Yaqoob}, T., \& {Wang}, J.~X. 2010, \apjs, 187, 581

\bibitem[{{Stern}(2015)}]{Stern:2015ve}
{Stern}, D. 2015, \apj, 807, 129

\bibitem[{{Stern} {et~al.}(2012){Stern}, {Assef}, {Benford}, {Blain}, {Cutri},
  {Dey}, {Eisenhardt}, {Griffith}, {Jarrett}, {Lake}, {Masci}, {Petty},
  {Stanford}, {Tsai}, {Wright}, {Yan}, {Harrison}, \& {Madsen}}]{Stern:2012fk}
{Stern}, D., {Assef}, R.~J., {Benford}, D.~J., {et~al.} 2012, \apj, 753, 30

\bibitem[{{Str{\"u}der} {et~al.}(2001){Str{\"u}der}, {Briel}, {Dennerl},
  {Hartmann}, {Kendziorra}, {Meidinger}, {Pfeffermann}, {Reppin}, {Aschenbach},
  {Bornemann}, {Br{\"a}uninger}, {Burkert}, {Elender}, {Freyberg}, {Haberl},
  {Hartner}, {Heuschmann}, {Hippmann}, {Kastelic}, {Kemmer}, {Kettenring},
  {Kink}, {Krause}, {M{\"u}ller}, {Oppitz}, {Pietsch}, {Popp}, {Predehl},
  {Read}, {Stephan}, {St{\"o}tter}, {Tr{\"u}mper}, {Holl}, {Kemmer}, {Soltau},
  {St{\"o}tter}, {Weber}, {Weichert}, {von Zanthier}, {Carathanassis}, {Lutz},
  {Richter}, {Solc}, {B{\"o}ttcher}, {Kuster}, {Staubert}, {Abbey}, {Holland},
  {Turner}, {Balasini}, {Bignami}, {La Palombara}, {Villa}, {Buttler},
  {Gianini}, {Lain{\'e}}, {Lumb}, \& {Dhez}}]{Struder:2001uq}
{Str{\"u}der}, L., {Briel}, U., {Dennerl}, K., {et~al.} 2001, \aap, 365, L18

\bibitem[{{Suganuma} {et~al.}(2006){Suganuma}, {Yoshii}, {Kobayashi},
  {Minezaki}, {Enya}, {Tomita}, {Aoki}, {Koshida}, \&
  {Peterson}}]{Suganuma:2006zr}
{Suganuma}, M., {Yoshii}, Y., {Kobayashi}, Y., {et~al.} 2006, \apj, 639, 46

\bibitem[{{Torricelli-Ciamponi} {et~al.}(2014){Torricelli-Ciamponi},
  {Pietrini}, {Risaliti}, \& {Salvati}}]{Torricelli-Ciamponi:2014dq}
{Torricelli-Ciamponi}, G., {Pietrini}, P., {Risaliti}, G., \& {Salvati}, M.
  2014, \mnras, 442, 2116

\bibitem[{{Tristram} \& {Schartmann}(2011)}]{Tristram:2011ve}
{Tristram}, K.~R.~W., \& {Schartmann}, M. 2011, \aap, 531, A99

\bibitem[{{Turner} {et~al.}(2001){Turner}, {Abbey}, {Arnaud}, {Balasini},
  {Barbera}, {Belsole}, {Bennie}, {Bernard}, {Bignami}, {Boer}, {Briel},
  {Butler}, {Cara}, {Chabaud}, {Cole}, {Collura}, {Conte}, {Cros}, {Denby},
  {Dhez}, {Di Coco}, {Dowson}, {Ferrando}, {Ghizzardi}, {Gianotti}, {Goodall},
  {Gretton}, {Griffiths}, {Hainaut}, {Hochedez}, {Holland}, {Jourdain},
  {Kendziorra}, {Lagostina}, {Laine}, {La Palombara}, {Lortholary}, {Lumb},
  {Marty}, {Molendi}, {Pigot}, {Poindron}, {Pounds}, {Reeves}, {Reppin},
  {Rothenflug}, {Salvetat}, {Sauvageot}, {Schmitt}, {Sembay}, {Short},
  {Spragg}, {Stephen}, {Str{\"u}der}, {Tiengo}, {Trifoglio}, {Tr{\"u}mper},
  {Vercellone}, {Vigroux}, {Villa}, {Ward}, {Whitehead}, \&
  {Zonca}}]{Turner:2001fk}
{Turner}, M.~J.~L., {Abbey}, A., {Arnaud}, M., {et~al.} 2001, \aap, 365, L27

\bibitem[{{Vaughan} {et~al.}(2003){Vaughan}, {Edelson}, {Warwick}, \&
  {Uttley}}]{Vaughan:2003uq}
{Vaughan}, S., {Edelson}, R., {Warwick}, R.~S., \& {Uttley}, P. 2003, \mnras,
  345, 1271

\bibitem[{{V{\'e}ron-Cetty} \& {V{\'e}ron}(2010)}]{Veron-Cetty:2010ly}
{V{\'e}ron-Cetty}, M.-P., \& {V{\'e}ron}, P. 2010, \aap, 518, A10

\bibitem[{{Walton} {et~al.}(2014){Walton}, {Risaliti}, {Harrison}, {Fabian},
  {Miller}, {Arevalo}, {Ballantyne}, {Boggs}, {Brenneman}, {Christensen},
  {Craig}, {Elvis}, {Fuerst}, {Gandhi}, {Grefenstette}, {Hailey}, {Kara},
  {Luo}, {Madsen}, {Marinucci}, {Matt}, {Parker}, {Reynolds}, {Rivers}, {Ross},
  {Stern}, \& {Zhang}}]{Walton:2014fk}
{Walton}, D.~J., {Risaliti}, G., {Harrison}, F.~A., {et~al.} 2014, \apj, 788,
  76

\bibitem[{{Wilms} {et~al.}(2000){Wilms}, {Allen}, \& {McCray}}]{Wilms:2000vn}
{Wilms}, J., {Allen}, A., \& {McCray}, R. 2000, \apj, 542, 914

\bibitem[{{Yaqoob}(2012)}]{Yaqoob:2012uq}
{Yaqoob}, T. 2012, \mnras, 423, 3360

\end{thebibliography}

\end{document}